# Bayesian linear regression with skew-symmetric error distributions with applications to survival analysis

Francisco J. Rubio[a*] and Marc G. Genton[b]

[a] University of Warwick, Department of Statistics, Coventry, CV4 7AL, UK.
[b] CEMSE Division, King Abdullah University of Science and Technology, Thuwal 23955-6900, Saudi Arabia.
* E-mail: Francisco.Rubio@warwick.ac.uk

## Abstract

We study Bayesian linear regression models with skew-symmetric scale mixtures of normal error distributions. These kinds of models can be used to capture departures from the usual assumption of normality of the errors in terms of heavy tails and asymmetry. We propose a general non-informative prior structure for these regression models and show that the corresponding posterior distribution is proper under mild conditions. We extend these propriety results to cases where the response variables are censored. The latter scenario is of interest in the context of accelerated failure time models, which are relevant in survival analysis. We present a simulation study that demonstrates good frequentist properties of the posterior credible intervals associated to the proposed priors. This study also sheds some light on the trade-off between increased model flexibility and the risk of over-fitting. We illustrate the performance of the proposed models with real data. Although we focus on models with univariate response variables, we also present some extensions to the multivariate case in the Supporting Web Material.

## 1 Introduction

The classical assumption of normality of the residual errors in linear regression models (LRMs) can be restrictive in practice. In many cases, the use of more flexible parametric distributional assumptions is necessary to capture departures from normality. These departures often arise when the sample contains outliers or when the residual errors are asymmetric. Several approaches have been proposed for parametrically modelling these departures from normality such as the use of scale mixtures of normal distributions [18], and skew-elliptical distributions [3, 34, 5], among others (see also [4] and [22]). In the absence of strong prior knowledge about the model parameters, a way for conducting Bayesian inference consists of using noninformative priors (often referred to as objective Bayesian inference). In a general sense, these kinds of priors are functions of the parameters that produce posterior distributions with good frequentist properties. The use of objective Bayesian inference in LRMs with symmetric errors has been widely studied. In particular, [18] studied the propriety of the posterior distribution associated with Bayesian LRMs with an improper prior structure and residual errors distributed according to the family of scale mixtures of normals (SMN). The use of more general prior structures has been recently studied in [13] and [24] for particular members of the SMN family. In a related vein, [36] studied the use of Jeffreys-type priors in the context of accelerated failure time (AFT) models with SMN errors (which are LRMs for the logarithm of the survival times). However, there are fewer studies related to the use of noninformative priors with flexible error distributions that allow for capturing skewness. [19] proposed a class of multivariate two-piece distributions that allow for capturing skewness. They employed these distributions for modelling the errors in multivariate LRMs and also proposed an improper prior structure, which is similar to that in [16]. Another recent reference is [30], who studied AFT models with errors distributed according to the generalised extreme value distribution. They proposed an improper prior for this LRM, but their conditions for the propriety of the posterior involve truncating the support of the prior distribution on the shape parameter, which in turns reduces the flexibility of the model, as well as a condition on the prior for the scale parameter.

Here, we study the use of the class of skew-symmetric scale mixtures of normal (SSSMN) distributions for modelling residual errors in LRMs. This class contains, for instance, the skew-normal distribution [2], the skew-$t$ distribution [3], and the skew-logistic distribution [37], as well as the corresponding symmetric models as particular cases. We propose a general improper prior structure that preserves the propriety of the posterior distribution in the sense that the posterior exists under the same conditions as in the model with SMN errors. This result allows us to appeal to existent results in order to show the propriety of the posterior associated to LRMs with SSSMN errors and the proposed prior structure. We discuss the use of the proposed Bayesian models in survival analysis as well as the impact of using flexible distributions in the prediction of the remaining life of patients that survived beyond the end of the study.

The paper is organised as follows. In Section 2 we describe the LRMs of interest and introduce the proposed noninformative prior structure. We provide easy to check sufficient conditions for the propriety of the corresponding posterior distribution. These results represent an extension to previous results for LRMs with symmetric errors [16, 18, 13, 24]. In Section 3, we extend our results to cases where the sample contains censored response variables for application of the proposed models to survival analysis. In Section 4, we present a simulation study where we illustrate the good frequentist properties of the 95% credible intervals obtained with the proposed priors and discuss some issues associated to certain skew-symmetric models. In Section 5 we present two applications with publicly available data sets in the context of biometric measures and survival analysis. All proofs are presented in the Supporting Web Material.

## 2 Linear regression with skew-symmetric errors

### 2.1 Bayesian model

Recall first that a real random variable $Z$ is said to be distributed according to a skew-symmetric distribution if its probability density function (PDF) can be written as [39]

$$s(z) = 2f(z)\varphi(z), \ \ z \in \mathbb{R}, \tag{1}$$

where $f$ is a symmetric PDF and $\varphi : \mathbb{R} \to [0,1]$ is a function that satisfies $\varphi(z) = 1 - \varphi(-z)$. We focus on the family of parametric skew-symmetric distributions of the type

$$s(z|\xi, \omega, \lambda, \delta) = \frac{2}{\omega} f\left(\left.\frac{z-\xi}{\omega}\right|\delta\right) G\left(\lambda \frac{z-\xi}{\omega}\right), \tag{2}$$

where $G$ is a cumulative distribution function (CDF) with continuous symmetric PDF $g$ with support on $\mathbb{R}$, $\xi \in \mathbb{R}$ is a location parameter, $\omega > 0$ is a scale parameter, $\lambda \in \mathbb{R}$, and $\delta \in \Delta \subset \mathbb{R}$ is a shape parameter. We will refer to $f$ as the "baseline density". Note that this structure allows for the case where $f = g$. We do not consider the case where $g$ contains an unknown shape parameter and $g \neq f$ since this would produce a model with 5 parameters that may play redundant roles. This structure covers many cases of practical interest such as the skew-normal distribution, the skew-$t$ distribution [3] with $\delta$ degrees of freedom, the skew-logistic distribution [37], and the skew-slash distribution [38]. Density (2) is asymmetric for $\lambda \neq 0$, and $s = f$ for $\lambda = 0$. If a random variable $Z$ has distribution (2), we will denote it as $Z \sim \text{SS}(\xi, \omega, \lambda, \delta; f, g)$.

Consider now the linear regression model:

$$y_j = \mathbf{x}_j^\top \boldsymbol{\beta} + \varepsilon_j, \ \ j = 1, \ldots, n, \tag{3}$$

where $y_j \in \mathbb{R}$, $\boldsymbol{\beta}$ is a $p$-dimensional vector of regression parameters, $\varepsilon_j \overset{i.i.d.}{\sim} \mathrm{SS}(0, \omega, \lambda, \delta; f, g)$, and $\mathbf{X} = (\mathbf{x}_1, \ldots, \mathbf{x}_n)^\top$ is a known $n \times p$ design matrix of full column rank. In some cases, we might be interested on centring the regression model on the mean or some quantile (such as the median), in which case we have to properly centre the error distribution. This centring strategy is typically done after the estimation of the parameters has been performed. This point has been discussed by [34] in the context of regression models with skew-elliptical distributions, as well as in [5].

We adopt the general prior structure:

$$\pi(\boldsymbol{\beta}, \omega, \lambda, \delta) = \frac{p(\lambda, \delta)}{\omega^a}, \qquad (4)$$

where $p(\lambda, \delta)$ is a proper prior. This prior structure covers the structure of several priors obtained by formal rules such as the independence Jeffreys prior, the Jeffreys-rule prior, and the reference prior [32]. Given that this prior is improper, we need to investigate conditions for the existence of the corresponding posterior distribution.

In order to provide a general propriety result, we restrict our study to the case when the baseline density $f$ in (2) belongs to the family of SMN distributions. Recall that a symmetric PDF $f$ is said to be a SMN if it can be written as:

$$f(z|\delta) = \int_{\mathbb{R}_+} \tau^{1/2} \phi(\tau^{1/2} z) dH(\tau|\delta), \quad z \in \mathbb{R},$$

where $H$ is a mixing distribution with positive support and $\phi$ represents the standard normal PDF. The family of scale mixtures of normals contains important models such as the normal distribution, the logistic distribution, the Student-$t$ distribution, the Laplace distribution, the symmetric generalised hyperbolic distribution, and the slash distribution. Under this setting, we have the following results (see the Supplementary Web Material for a formal proof).

(i) Consider the model (3)–(4). Assume that $f$ in (2) is a scale mixture of normals and that $\mathrm{rank}(\mathbf{X}) = p$. If $a = 1$, a sufficient condition for the propriety of the posterior of $(\boldsymbol{\beta}, \omega, \delta, \lambda)$ is $n > p$.

(ii) If $a > 1$, a sufficient condition for the propriety of the posterior is $n > p + 1 - a$ and

$$\int \tau^{-\frac{a-1}{2}} \tilde{p}(\delta) dH(\tau|\delta) d\delta < \infty, \qquad (5)$$

where $\tilde{p}(\delta)$ represents the marginal prior on $\delta$.

These results indicate that the introduction of skewness in the distribution of the residual errors, by means of the skew-symmetric construction (2), does not affect the existence of the posterior distribution, as long as we use a proper prior on the skewness parameter $\lambda$. For $a = 1$ the propriety of the posterior is guaranteed by having more observations than covariates for the class of LRMs with SSSMN errors. The results presented here are satisfied with probability one (for almost all sets of observations). We refer the reader to [16, 17], [18], and [36] for a discussion on zero-probability events that induce improper posteriors. This includes, for instance, samples with a certain number of responses that can be represented as an exact combination of their covariates which, in principle, have probability zero of occurrence under a continuous model but that may appear in practice due to rounding of the measurements. For $a > 1$ the conditions for the existence of the posterior distribution become more restrictive and condition (5) has to be checked case by case. For the cases when the residual error distribution is a skew-normal or a skew-logistic (normal or logistic baseline $f$), the condition (5) is automatically satisfied [36], and, therefore, the condition $n > p + 1 - a$ is sufficient for the propriety of the posterior. Similarly, for cases when the baseline PDF $f$ is a generalised hyperbolic distribution (with either fixed tail parameter $\delta > 0$ or with a compactly supported marginal prior $\tilde{p}(\delta)$) or a Laplace distribution, the condition $n > p + 1 - a$ is also sufficient for the existence of the posterior [13, 24].

### 2.1.1 The role of $\lambda$ in skew-symmetric models

An aspect that has been little discussed in the literature is the overall influence of the parameter $\lambda$ on the shape of a skew-symmetric PDF (1). We already mentioned that the PDF (1) is left/right skewed for $\lambda \lessgtr 0$. However, the parameter $\lambda$ controls other features as well such as spread and location of the mode. Moreover, for some combinations of $f$ and $G$, the parameter $\lambda$ has little influence of the shape of the PDF for a certain range of values around $\lambda = 0$. For instance, in the skew-normal PDF, the shape parameter $\lambda$ has little influence on the asymmetry when $|\lambda| \leq 1$. In this region, $\lambda$ mainly controls the location and spread of the PDF. Figure 1 shows that a skew-normal PDF with parameter $\lambda = 1$ can be reasonably well approximated with a normal PDF. This phenomenon is also observed in other models where $f$ and $G$ have lighter tails than normal, while models with heavier tails seem to be exempt from this problem. Intuitively, this suggests that it is hard to estimate the parameters $(\xi, \omega, \lambda)$ (both from a Classical and a Bayesian perspective) in some light-tailed skew-symmetric models when the true value lies in a certain interval centred at $\lambda = 0$ and the sample size is small or moderate. Consequently, if one uses a skew-symmetric distribution, with lighter or equal tails than normal, to model the errors in a LRM, it is expected to observe high correlation between the estimator of the intercept regression parameter, the scale parameter, and the skewness parameter (See Section 4). Given that these issues only appear when the errors are nearly symmetric, a simple solution consists of testing for $\lambda = 0$ in order to identify the best and most parsimonious model for the data. Moreover, in the context of our paper this issue has little relevance since we are only considering the cases when $f$ is a SMN, leaving the normal as the only problematic case.

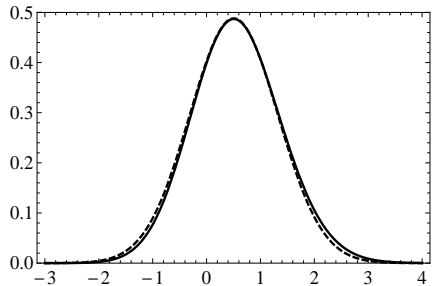

Figure 1: Normal PDF with parameters $(0.505, 0.817)$ (dashed line) *vs.* skew-normal PDF with parameters $(0, 1, 1)$ (continuous line).

## 2.2 Discussion on the choice of the priors for the shape parameters

The choice of the priors for the shape parameters $(\lambda, \delta)$ deserves further discussion. We present a summary of some informative an non-informative priors used for the shape parameters in skew-symmetric models.

### 2.2.1 Priors for $\lambda$

(i) **Jeffreys priors**. For the skew-normal regression model, we can employ the reference prior of the skewness parameter $\lambda$. This prior was obtained by [26], who showed that this is proper and well defined. [8] showed that this prior can be reasonably well approximated using a Student-$t$ distribution with $1/2$ degrees of freedom and scale parameter $\sigma_0 = \pi/2$. In [9], a similar result was also established for the skew-$t$ model with degrees of freedom $\delta > 2$. [32] characterised the marginal reference prior of the parameter $\lambda$ in a large class of skew-symmetric models and showed that this is symmetric, with tails of order $O(|\lambda|^{-3/2})$, and proper. These kinds of priors have been shown to induce posteriors with good frequentist properties [26, 9, 32]. However, the reference priors of $\lambda$ are well defined at $\lambda = 0$ only if the second moment of $f$ is finite. As shown in [32], the Jeffreys/reference prior of $\lambda$ can be calculated as follows

$$p(\lambda) \propto \sqrt{\int_0^\infty x^2 f(x) \frac{g(\lambda x)^2}{G(\lambda x)[1 - G(\lambda x)]} dx}. \tag{6}$$

(ii) **Matching priors**. [10] calculated the matching prior for the skewness parameter $\lambda$ in the skew-normal model. They showed that this prior is proper, data-independent, bimodal, with tails of order $O(|\lambda|^{-3/2})$, and that this prior induces a posterior with good frequentist properties. However, the bimodality of this prior might complicate sampling from the posterior for small and moderate samples. To the best of our knowledge, matching priors have not been calculated for other skew-symmetric models.

(iii) **Heuristic priors**. An alternative non-informative prior can be constructed by using the popular alternative parameterisation of the skew-normal distribution in terms of the parameter $\rho = \lambda/\sqrt{1 + \lambda^2} \in (-1, 1)$. If we assign a uniform prior on the parameter $\rho$, this induces the following proper prior on $\lambda$:

$$p(\lambda) \propto \frac{1}{(1 + \lambda^2)^{3/2}}. \tag{7}$$

This prior has been used recently in [11].

(iv) **Informative priors**. In cases where there is reliable prior information, one can appeal to any proper prior that can capture the features of our prior beliefs. In particular, [11] proposed the use of the skew-normal distribution as a prior distribution, with hyperparameters $(\xi_0, \omega_0, \lambda_0)$, for $\lambda$. This prior allows the user to control the mass allocated on values of $\lambda \lessgtr 0$, which are related to negative and positive skewness.

### 2.2.2 Priors for $\delta$

Regarding the tail parameter $\delta$, [33] proposed a general method for constructing weakly informative priors for kurtosis parameters. The idea consists of assigning a uniform prior to a bounded measure of kurtosis applied to the symmetric baseline density $f(\cdot|\delta)$. In order for this method to be applicable, the chosen measure of kurtosis must be a one-to-one function of the parameter $\delta$. This strategy induces a proper prior on the parameter $\delta$ which can be interpreted as a weakly informative prior, in the sense that it assigns a flat prior on a function that represents the influence of the parameter $\delta$ on the shape of the density. In the case of the degrees of freedom of the Student-$t$ distribution, [33] showed that this strategy produces a prior with a behaviour similar to that of the approximation to the Jeffreys prior for this parameter [21] proposed in [23].

In addition, these priors on $\delta$ can be coupled with the Jeffreys priors in order to produce a joint prior on $(\lambda, \delta)$ by using the decomposition $p(\lambda, \delta) = p(\lambda|\delta)p(\delta)$, where

$$p(\lambda|\delta) \propto \sqrt{\int_0^\infty x^2 f(x|\delta) \frac{g(\lambda x)^2}{G(\lambda x)[1 - G(\lambda x)]} dx}.$$

This prior also has tails of order $O(|\lambda|^{-3/2})$, for each value of $\delta$. If the prior $p(\lambda|\delta)$ is not too variable for different values of $\delta$, we may opt for an independent structure $p(\lambda, \delta) = p(\lambda)p(\delta)$ for practical purposes. For instance, in the skew-$t$ distribution, the prior $p(\lambda|\delta)$ can be reasonably well approximated with a Student-$t$ distribution with $1/2$ degrees of freedom and scale parameter $\sigma_0$. A reasonable value of $\sigma_0$ depends on the degrees of freedom $\delta$. Figure 2 shows the values of $\sigma_0$ that produce a reasonable approximation (obtained by matching the mode) for $\delta = 2.1, 2.2, \ldots, 50$. We observe that $\sigma_0$ varies between 0.5 and $\pi/2$. In this case, for the sake of simplicity, we adopt the prior $p(\lambda, \delta) = p(\lambda)p(\delta)$, where $p(\lambda)$ is a Student-$t$ distribution with $1/2$ degrees of freedom and scale parameter $\sigma_0 = \pi/2$, which corresponds to the reference prior of $\lambda$. In Section 4 we show, through a simulation study, that this prior structure induces a posterior with good frequentist properties.

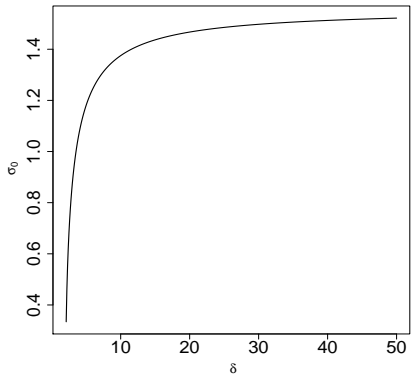

Figure 2: Value of the scale parameter $\omega_0$ used in the Student-$t$ approximation to the Jeffreys prior of $\lambda$ in the skew-$t$ model with $\delta$ degrees of freedom.

## 3 Accelerated failure time models

A closely related family of LRMs that are of great interest in survival analysis are AFT models. The basic idea behind this kind of LRM consists of modelling a set of survival times $\mathbf{T} = (T_1, \ldots, T_n)^\top$ in terms of a set of covariates $\boldsymbol{\beta}$ through the model equation:

$$\log T_j = \mathbf{x}_j^\top \boldsymbol{\beta} + \varepsilon_j, \ j = 1, \ldots, n. \tag{8}$$

In this context, the distribution of the errors $\varepsilon_j$ is typically assumed to be normal or that $\exp(\varepsilon_j)$ follows a Weibull distribution. Given that the normality assumption can be restrictive in practice, several alternative models have been proposed. For instance, from a classical inferential framework, [7] proposed the use of the Birnbaum-Saunders distribution, which corresponds to modelling the survival times $T_j$ as a log-Birnbaum-Saunders distribution; and [31] employed the class of log-two-piece distributions. From a Bayesian perspective, [36] employed the class of log scale mixture of normals for modelling survival times; meanwhile [30] employed the log generalised extreme value distribution.

We consider the class of log skew-symmetric (LSS) distributions [28]:

$$s_l(t|\xi, \omega, \lambda, \delta) = \frac{2}{\omega t} f\left(\frac{\log t - \xi}{\omega}\bigg|\delta\right) G\left(\lambda \frac{\log t - \xi}{\omega}\right), \ t > 0, \tag{9}$$

for modelling the survival times $T_j$ in (8), where $f$ and $g$ are as described in Section 2. It includes, for example, the log-skew-normal and log-skew-$t$ distributions [6, 25]. This class of distributions can also be seen as an extension of the class of log-symmetric distributions (which is obtained with $\lambda = 0$). If a positive random variable $T$ has distribution (9) we denote it as $T \sim \text{LSS}(\xi, \omega, \lambda, \delta; f, g)$. More specifically, we assume that $T_j|\mathbf{x}_j, \boldsymbol{\beta}, \omega, \lambda, \delta \sim \text{LSS}(\mathbf{x}_j^\top \boldsymbol{\beta}, \omega, \lambda, \delta; f, g)$, that $f$ in (9) is a scale mixture of normals, and that $\text{rank}(\mathbf{X}) = p$. If we adopt the prior structure (4) for this model, then the corresponding posterior is proper under the conditions in Section 2. However, in the context of AFT models, the presence of (right–, left–, and interval–) censored responses is quite common. Suppose that $T_j|\mathbf{x}_j, \boldsymbol{\beta}, \omega, \lambda, \delta \sim \text{LSS}(\mathbf{x}_j^\top \boldsymbol{\beta}, \omega, \lambda, \delta; f, g)$, that $f$ in (9) is a scale mixture of normals, and that $\text{rank}(\mathbf{X}) = p$. Consider the prior structure (4) for this model. Suppose also that $n_c \leq n$ observations are censored and $n_o = n - n_c$ are observed. Then,

(i) If $a = 1$, a sufficient condition for the propriety of the corresponding posterior is $n_o > p$.

(ii) If $a > 1$, a sufficient condition for the propriety of the posterior is $n_o > p + 1 - a$ together with (5).

In the Supplementary Web Material we provide a study of the extreme case in which all of the observations are censored (Corollary 2). This scenario requires a more careful analysis given that, intuitively, samples containing only censored observations contain little information about the parameters, thus, one has to be careful about using non-informative priors in such extreme scenarios.

## 4 Simulation study

In this section we present a simulation study that illustrates the performance of the proposed models. We study the regression model:

$$y_j = \beta_1 + \beta_2 x_{1j} + \beta_3 x_{2j} + \varepsilon_j, \ \ j = 1, \ldots, n,$$

where we simulate the variables $x_{1j}$ and $x_{2j}$ from a standard normal distribution and consider different combinations of the distribution of the residual errors and the sample size $n$.

In the first scenario, we simulate the residual errors from a skew-normal distribution with scale parameter $\omega = 0.5$ and skewness parameter $\lambda = 0, 1, 2, 3, 4, 5$, $(\beta_1, \beta_2, \beta_3) = (1, 2, 3)$ and $n = 100, 250, 500$. Negative values of $\lambda$ produce similar results and are, therefore, omitted. We adopt the product prior structure (4) with $a = 1$ and $p(\lambda)$ given by the reference prior of this parameter [26]. We use the Student-$t$ approximation from [8] to facilitate its implementation. For each of these scenarios, we obtain $N = 1,000$ samples of size $2,000$ from the posterior distribution using the R [29] t-walk Markov Chain Monte Carlo (MCMC) sampler [14] after a burn-in period of $5,000$ iterations and thinned to every 25th iteration (55,000 iterations in total). Then, we calculate the proportion of 95% credible intervals that include the true value of the parameters, the 5%, 50%, and 95% quantiles of the maximum likelihood estimators (MLEs), the maximum *a posteriori* (MAP) estimators, and posterior median estimators, as well as the median Bayes factor associated to the hypothesis $H_0 : \lambda = 0$. The Bayes factor is approximated using the Savage-Dickey density ratio [15]. In the second scenario, we simulate the residual error from a skew-logistic distribution with scale parameter $\omega = 0.5$ and skewness parameter $\lambda = 0, 1, 2, 3, 4, 5$. We adopt the product prior structure (4) with $a = 1$ and $p(\lambda)$ given by the reference prior of this parameter [32]. We use the Student-$t$ approximation (1/2 degrees of freedom and scale 4/3) from [32] to facilitate its implementation. In the third scenario, we simulate the residual errors from a skew-$t$ distribution with degrees of freedom $\delta = 3$ and scale parameter $\omega = 0.5$. We adopt the prior on the skewness parameter described in the first simulation scenario. For the degrees of freedom $\delta$, we use the approximation to the Jeffreys prior proposed in [23]:

$$\pi(\delta) = \frac{2d\delta}{(\delta + d)^3}. \tag{10}$$

We fix the hyperparameter $d = 2$, which induces a prior with mode at $\delta = 1$.

Results are reported in Tables 1–3 below and Section 3, Tables 1S–6S, of the Supplementary Web Material. For the first scenario (skew-normal errors, Tables 1–3), we can observe that the coverage of the credible intervals of the parameter $\omega$ is greatly affected when the true value of $\lambda$ is zero or one. Moreover, the Bayes factors associated to $H_0 : \lambda = 0$ favours the normal model in both cases, for all sample sizes, when $\lambda = 0, 1$, which emphasises the difficulty in identifying models with $|\lambda| \leq 1$. These scenarios correspond to the cases described in Section 2.1.1 where the skew-normal distribution is nearly or exactly symmetric, which in turns complicates the identification of the parameters $(\beta_1, \omega, \lambda)$, and induces a strong correlation between these parameters. Figures 3–4 show the scatter plots associated to a posterior sample of $(\beta_1, \beta_2, \beta_3, \omega, \lambda)$ for $n = 250$ and $\lambda = 1, 5$, respectively. These figures show the strong correlation between the parameters $(\beta_1, \omega, \lambda)$ for $\lambda = 1$, while the correlation between these parameters seems to be lower for $\lambda = 5$. We can also observe from these tables that the coverage improves as $\lambda \geq 2$ and that the coverage proportions converge to the nominal value as the sample size increases. This study also indicates that we need at least 100 observations in order to get a good coverage. For the second and third scenarios (Tables 1S–6S), we observe a good coverage for all the sample sizes as well as a good behaviour of the Bayesian estimators compared to that of the corresponding MLEs.

In the Supplementary Web Material, Tables 7S–9S, we present an additional simulation study using the configuration in the first scenario with the prior (7) for the parameter $\lambda$. This study shows that the prior (7), that has been considered as a noninformative prior in the literature, induces a posterior with poor frequentist properties.

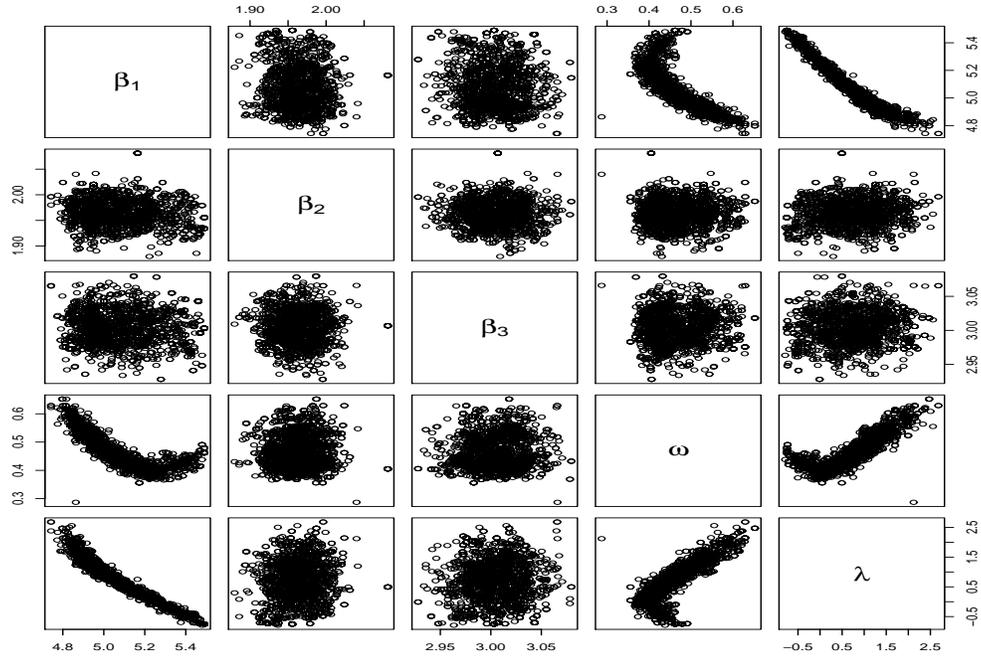

Figure 3: Scatter plot for a single posterior simulation of $(\beta_1, \beta_2, \beta_3, \omega, \lambda)$ with $n = 250$ and $\lambda = 1$.

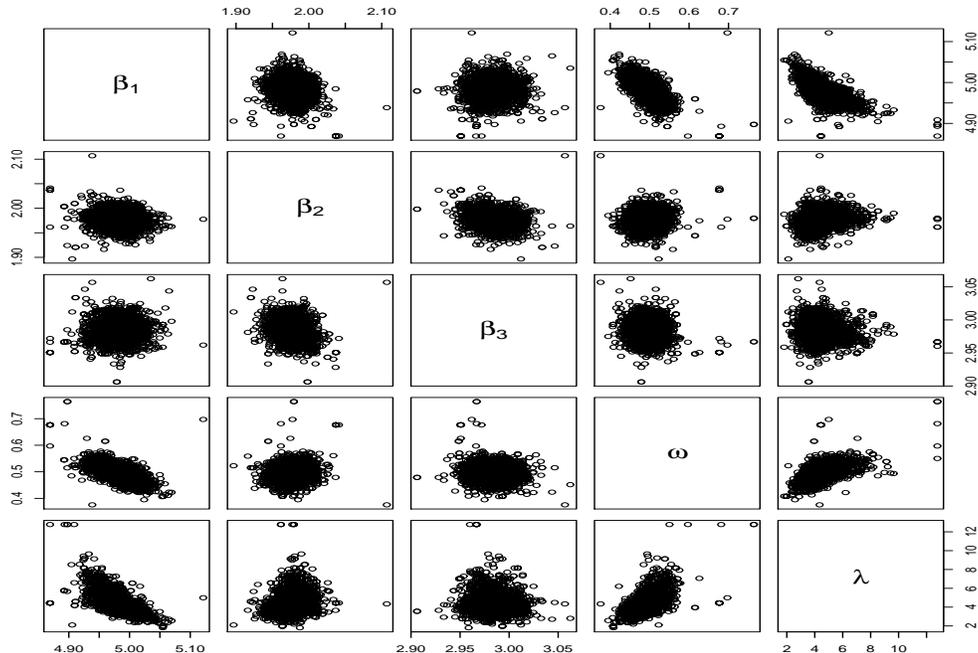

Figure 4: Scatter plot for a single posterior simulation of $(\beta_1, \beta_2, \beta_3, \omega, \lambda)$ with $n = 250$ and $\lambda = 5$.

## 5 Applications

In this section, we illustrate the performance of the proposed Bayesian models with two real data sets. In the first example, we revisit the popular "Australian Athletes" data set. We compare the performance of 4 types of distributional assumptions on the residual errors in a LRM studied in [1]. In the second example we present an application in the context of survival analysis of cancer patients. Simulations from the corresponding posterior distributions are obtained using the t–walk [14]. Model comparison is conducted via Bayes factors which are calculated using an importance sampling (and corroborated using Laplace's method), the Bayesian information criterion (BIC), and log-predictive marginal likelihood (LPML). LPML is a measure that ranks the models of interest in terms of their predictive performance (see [35] for an overview that includes the case with interval censored observations). Larger values of LPML indicate a better predictive performance. The use of the Bayes factors with improper priors is justified in this context since we use the same prior structure (with improper priors only on the common parameters) for the different models in order to avoid the presence of arbitrary constants (see [36] for a discussion on this point). R codes are available upon request.

### 5.1 Australian athletes data

Our first application concerns the study of the regression model [1]:

$$Lbm_j = \beta_1 Ht_j + \beta_2 Wt_j + \varepsilon_j, \ \ j = 1, \ldots, 102, \tag{11}$$

where $Lbm_j$, $Ht_j$, and $Wt_j$ denote the lean body mass, the height, and the weight of 102 Australian male athletes from the Australian Institute of Sport (AIS) data set. We compare the models obtained with four distributional assumptions on the residual errors $\varepsilon_j$: a skew-normal distribution $SN(0, \omega, \lambda)$ [2], a skew Student-$t$ distribution $SSt(0, \omega, \lambda, \delta)$ with unknown degrees of freedom, and the corresponding symmetric submodels (normal and Student-$t$). We adopt the prior structure (4) with $a = 1$. For the

| Method | MLE | | | MAP | | | Median | | | Coverage | BF |
|---|---|---|---|---|---|---|---|---|---|---|---|
| | 5% | 50% | 95% | 5% | 50% | 95% | 5% | 50% | 95% | | |
| $\lambda = 0$ | | | | | | | | | | | |
| $\beta_1$ | 4.456 | 4.917 | 5.521 | 4.463 | 4.996 | 5.519 | 4.559 | 4.999 | 5.380 | 0.987 | – |
| $\beta_2$ | 1.913 | 2.002 | 2.089 | 1.911 | 2.001 | 2.091 | 1.916 | 2.003 | 2.085 | 0.956 | – |
| $\beta_3$ | 2.908 | 3.001 | 3.087 | 2.905 | 2.999 | 3.086 | 2.906 | 3.000 | 3.087 | 0.942 | – |
| $\omega$ | 0.492 | 0.613 | 0.761 | 0.475 | 0.542 | 0.714 | 0.504 | 0.581 | 0.699 | 0.849 | – |
| $\lambda$ | -1.993 | 0.285 | 2.224 | -1.452 | -0.003 | 1.613 | -1.196 | 0.009 | 1.520 | 0.989 | 2.141 |
| $\lambda = 1$ | | | | | | | | | | | |
| $\beta_1$ | 4.788 | 5.009 | 5.317 | 4.796 | 5.057 | 5.649 | 4.856 | 5.193 | 5.507 | 0.947 | – |
| $\beta_2$ | 1.931 | 2.001 | 2.069 | 1.927 | 2.001 | 2.069 | 1.928 | 2.002 | 2.067 | 0.950 | – |
| $\beta_3$ | 2.930 | 3.001 | 3.066 | 2.929 | 3.001 | 3.071 | 2.931 | 3.002 | 3.066 | 0.963 | – |
| $\omega$ | 0.373 | 0.485 | 0.635 | 0.388 | 0.450 | 0.613 | 0.417 | 0.483 | 0.603 | 0.994 | – |
| $\lambda$ | 0.007 | 0.992 | 2.550 | -0.988 | 0.190 | 2.063 | -0.713 | 0.248 | 1.974 | 0.964 | 2.029 |
| $\lambda = 2$ | | | | | | | | | | | |
| $\beta_1$ | 4.886 | 5.003 | 5.252 | 4.885 | 5.009 | 5.435 | 4.900 | 5.059 | 5.359 | 0.911 | – |
| $\beta_2$ | 1.941 | 2.001 | 2.058 | 1.938 | 2.003 | 2.060 | 1.942 | 2.001 | 2.058 | 0.955 | – |
| $\beta_3$ | 2.944 | 3.000 | 3.060 | 2.941 | 3.001 | 3.061 | 2.943 | 3.001 | 3.059 | 0.952 | – |
| $\omega$ | 0.354 | 0.496 | 0.600 | 0.336 | 0.447 | 0.593 | 0.365 | 0.462 | 0.589 | 0.947 | – |
| $\lambda$ | 0.451 | 2.083 | 4.942 | -0.315 | 1.585 | 3.389 | 0.035 | 1.441 | 3.973 | 0.901 | 1.093 |
| $\lambda = 3$ | | | | | | | | | | | |
| $\beta_1$ | 4.909 | 5.002 | 5.115 | 4.904 | 5.003 | 5.147 | 4.915 | 5.020 | 5.257 | 0.912 | – |
| $\beta_2$ | 1.946 | 1.998 | 2.047 | 1.944 | 1.998 | 2.050 | 1.947 | 1.999 | 2.047 | 0.955 | – |
| $\beta_3$ | 2.949 | 3.000 | 3.054 | 2.948 | 3.000 | 3.054 | 2.950 | 3.001 | 3.053 | 0.950 | – |
| $\omega$ | 0.395 | 0.491 | 0.590 | 0.319 | 0.484 | 0.585 | 0.359 | 0.482 | 0.587 | 0.929 | – |
| $\lambda$ | 1.488 | 3.225 | 10.269 | -0.154 | 2.463 | 5.112 | 0.577 | 2.742 | 6.939 | 0.904 | 0.223 |
| $\lambda = 4$ | | | | | | | | | | | |
| $\beta_1$ | 4.928 | 5.002 | 5.084 | 4.922 | 5.000 | 5.091 | 4.929 | 5.008 | 5.137 | 0.946 | – |
| $\beta_2$ | 1.954 | 2.000 | 2.046 | 1.953 | 2.001 | 2.047 | 1.954 | 2.001 | 2.045 | 0.964 | – |
| $\beta_3$ | 2.954 | 3.002 | 3.047 | 2.953 | 3.003 | 3.048 | 2.955 | 3.002 | 3.048 | 0.964 | – |
| $\omega$ | 0.407 | 0.495 | 0.576 | 0.369 | 0.488 | 0.574 | 0.380 | 0.490 | 0.574 | 0.950 | – |
| $\lambda$ | 2.298 | 4.505 | 108.781 | -1.076 | 3.227 | 8.417 | 1.502 | 3.807 | 12.553 | 0.941 | $5 \times 10^{-2}$ |
| $\lambda = 5$ | | | | | | | | | | | |
| $\beta_1$ | 4.933 | 4.998 | 5.068 | 4.930 | 4.997 | 5.074 | 4.938 | 5.000 | 5.096 | 0.954 | – |
| $\beta_2$ | 1.954 | 1.998 | 2.043 | 1.953 | 1.999 | 2.043 | 1.956 | 1.999 | 2.044 | 0.955 | – |
| $\beta_3$ | 2.960 | 3.003 | 3.046 | 2.959 | 3.003 | 3.047 | 2.962 | 3.003 | 3.045 | 0.969 | – |
| $\omega$ | 0.418 | 0.499 | 0.575 | 0.397 | 0.494 | 0.574 | 0.402 | 0.498 | 0.578 | 0.958 | – |
| $\lambda$ | 2.997 | 5.883 | 138.560 | -17.640 | 3.930 | 12.810 | 2.192 | 4.929 | 21.107 | 0.943 | $2 \times 10^{-2}$ |

Table 1: Point estimators, coverage proportions and Bayes factors: Linear regression model with residual errors simulated from a skew-normal distribution, $n = 100$.

| Method | MLE | | | MAP | | | Median | | | Coverage | BF |
|---|---|---|---|---|---|---|---|---|---|---|---|
| | 5% | 50% | 95% | 5% | 50% | 95% | 5% | 50% | 95% | | |
| $\lambda = 0$ | | | | | | | | | | | |
| $\beta_1$ | 4.574 | 4.965 | 5.424 | 4.568 | 5.008 | 5.429 | 4.665 | 4.994 | 5.335 | 0.988 | – |
| $\beta_2$ | 1.948 | 2.000 | 2.057 | 1.945 | 1.999 | 2.057 | 1.947 | 2.000 | 2.056 | 0.945 | – |
| $\beta_3$ | 2.947 | 2.998 | 3.052 | 2.945 | 2.996 | 3.055 | 2.946 | 2.997 | 3.053 | 0.952 | – |
| $\omega$ | 0.504 | 0.584 | 0.683 | 0.484 | 0.525 | 0.656 | 0.508 | 0.555 | 0.640 | 0.859 | – |
| $\lambda$ | -1.406 | 0.079 | 1.400 | -1.240 | 0.030 | 1.234 | -1.035 | 0.015 | 1.017 | 0.988 | 2.547 |
| $\lambda = 1$ | | | | | | | | | | | |
| $\beta_1$ | 4.869 | 5.000 | 5.290 | 4.872 | 5.024 | 5.548 | 4.897 | 5.157 | 5.430 | 0.903 | – |
| $\beta_2$ | 1.957 | 2.000 | 2.047 | 1.957 | 2.000 | 2.049 | 1.959 | 2.000 | 2.047 | 0.943 | – |
| $\beta_3$ | 2.957 | 3.001 | 3.044 | 2.956 | 3.001 | 3.044 | 2.958 | 3.001 | 3.043 | 0.949 | – |
| $\omega$ | 0.393 | 0.498 | 0.589 | 0.401 | 0.440 | 0.581 | 0.420 | 0.465 | 0.564 | 0.986 | – |
| $\lambda$ | 0.014 | 1.018 | 1.782 | -0.809 | 0.548 | 1.690 | -0.428 | 0.390 | 1.553 | 0.903 | 2.282 |
| $\lambda = 2$ | | | | | | | | | | | |
| $\beta_1$ | 4.922 | 5.002 | 5.100 | 4.922 | 5.002 | 5.117 | 4.928 | 5.013 | 5.217 | 0.919 | – |
| $\beta_2$ | 1.967 | 2.001 | 2.035 | 1.966 | 2.001 | 2.038 | 1.966 | 2.001 | 2.035 | 0.955 | – |
| $\beta_4$ | 2.966 | 3.000 | 3.033 | 2.965 | 3.001 | 3.034 | 2.966 | 3.001 | 3.033 | 0.967 | – |
| $\omega$ | 0.420 | 0.498 | 0.565 | 0.358 | 0.492 | 0.564 | 0.386 | 0.489 | 0.564 | 0.919 | – |
| $\lambda$ | 1.231 | 2.046 | 3.159 | 0.718 | 1.871 | 2.886 | 0.592 | 1.897 | 3.009 | 0.917 | 0.080 |
| $\lambda = 3$ | | | | | | | | | | | |
| $\beta_1$ | 4.948 | 5.001 | 5.064 | 4.946 | 5.001 | 5.067 | 4.949 | 5.005 | 5.074 | 0.944 | – |
| $\beta_2$ | 1.971 | 2.000 | 2.028 | 1.970 | 2.001 | 2.029 | 1.971 | 2.000 | 2.028 | 0.966 | – |
| $\beta_3$ | 2.970 | 3.001 | 3.029 | 2.969 | 3.000 | 3.029 | 2.970 | 3.000 | 3.029 | 0.963 | – |
| $\omega$ | 0.437 | 0.497 | 0.552 | 0.431 | 0.494 | 0.550 | 0.430 | 0.495 | 0.551 | 0.950 | – |
| $\lambda$ | 2.034 | 3.093 | 4.676 | 1.851 | 2.845 | 4.117 | 1.826 | 2.950 | 4.330 | 0.947 | $1 \times 10^{-11}$ |
| $\lambda = 4$ | | | | | | | | | | | |
| $\beta_1$ | 4.958 | 4.998 | 5.050 | 4.958 | 4.998 | 5.051 | 4.960 | 5.000 | 5.054 | 0.962 | – |
| $\beta_2$ | 1.973 | 2.000 | 2.028 | 1.972 | 2.000 | 2.027 | 1.973 | 2.000 | 2.027 | 0.965 | – |
| $\beta_3$ | 2.973 | 2.999 | 3.026 | 2.973 | 2.999 | 3.027 | 2.973 | 2.999 | 3.026 | 0.970 | – |
| $\omega$ | 0.449 | 0.497 | 0.549 | 0.445 | 0.495 | 0.545 | 0.446 | 0.496 | 0.546 | 0.966 | – |
| $\lambda$ | 2.820 | 4.180 | 6.669 | 2.539 | 3.781 | 5.629 | 2.687 | 3.979 | 6.161 | 0.947 | $1 \times 10^{-13}$ |
| $\lambda = 5$ | | | | | | | | | | | |
| $\beta_1$ | 4.962 | 4.997 | 5.042 | 4.960 | 4.997 | 5.044 | 4.962 | 4.999 | 5.044 | 0.953 | – |
| $\beta_2$ | 1.976 | 2.001 | 2.025 | 1.975 | 2.001 | 2.026 | 1.976 | 2.001 | 2.026 | 0.967 | – |
| $\beta_3$ | 2.975 | 3.000 | 3.026 | 2.974 | 2.999 | 3.026 | 2.974 | 2.999 | 3.026 | 0.959 | – |
| $\omega$ | 0.447 | 0.499 | 0.548 | 0.447 | 0.497 | 0.545 | 0.448 | 0.499 | 0.547 | 0.956 | – |
| $\lambda$ | 3.550 | 5.289 | 8.647 | 3.196 | 4.692 | 7.200 | 3.356 | 4.965 | 7.821 | 0.959 | $6 \times 10^{-12}$ |

Table 2: Point estimators, coverage proportions and Bayes factors: Linear regression model with residual errors simulated from a skew-normal distribution, $n = 250$.

| Method | MLE | | | MAP | | | Median | | | Coverage | BF |
|---|---|---|---|---|---|---|---|---|---|---|---|
| | 5% | 50% | 95% | 5% | 50% | 95% | 5% | 50% | 95% | | |
| $\lambda = 0$ | | | | | | | | | | | |
| $\beta_1$ | 4.612 | 5.022 | 5.376 | 4.608 | 4.991 | 5.379 | 4.680 | 5.000 | 5.287 | 0.981 | – |
| $\beta_2$ | 1.964 | 2.001 | 2.035 | 1.963 | 2.000 | 2.037 | 1.965 | 2.001 | 2.036 | 0.957 | – |
| $\beta_3$ | 2.962 | 3.000 | 3.039 | 2.959 | 3.000 | 3.040 | 2.962 | 3.000 | 3.040 | 0.951 | – |
| $\omega$ | 0.502 | 0.567 | 0.649 | 0.487 | 0.520 | 0.630 | 0.507 | 0.541 | 0.615 | 0.848 | – |
| $\lambda$ | -1.134 | -0.045 | 1.181 | -1.066 | 0.009 | 1.140 | -0.807 | -0.007 | 0.898 | 0.982 | 2.953 |
| $\lambda = 1$ | | | | | | | | | | | |
| $\beta_1$ | 4.897 | 5.000 | 5.263 | 4.896 | 5.012 | 5.450 | 4.910 | 5.112 | 5.333 | 0.892 | – |
| $\beta_2$ | 1.969 | 1.999 | 2.029 | 1.967 | 1.999 | 2.031 | 1.969 | 1.999 | 2.031 | 0.948 | – |
| $\beta_3$ | 2.970 | 3.000 | 3.031 | 2.968 | 3.000 | 3.033 | 2.970 | 3.000 | 3.032 | 0.950 | – |
| $\omega$ | 0.411 | 0.501 | 0.569 | 0.405 | 0.437 | 0.567 | 0.424 | 0.463 | 0.560 | 0.965 | – |
| $\lambda$ | 0.055 | 1.010 | 1.602 | -0.430 | 0.833 | 1.547 | -0.130 | 0.544 | 1.505 | 0.885 | 2.279 |
| $\lambda = 2$ | | | | | | | | | | | |
| $\beta_1$ | 4.950 | 4.999 | 5.062 | 4.950 | 5.000 | 5.064 | 4.954 | 5.004 | 5.081 | 0.950 | – |
| $\beta_2$ | 1.975 | 2.001 | 2.025 | 1.974 | 2.000 | 2.025 | 1.975 | 2.000 | 2.024 | 0.953 | – |
| $\beta_3$ | 2.976 | 3.001 | 3.026 | 2.976 | 3.001 | 3.027 | 2.976 | 3.001 | 3.026 | 0.949 | – |
| $\omega$ | 0.447 | 0.499 | 0.545 | 0.440 | 0.497 | 0.544 | 0.436 | 0.496 | 0.543 | 0.942 | – |
| $\lambda$ | 1.446 | 2.022 | 2.706 | 1.362 | 1.943 | 2.585 | 1.317 | 1.961 | 2.633 | 0.937 | $2\times10^{-26}$ |
| $\lambda = 3$ | | | | | | | | | | | |
| $\beta_1$ | 4.963 | 4.999 | 5.040 | 4.962 | 5.000 | 5.041 | 4.965 | 5.000 | 5.043 | 0.958 | – |
| $\beta_2$ | 1.979 | 1.999 | 2.020 | 1.978 | 1.999 | 2.022 | 1.979 | 1.999 | 2.021 | 0.967 | – |
| $\beta_3$ | 2.977 | 3.000 | 3.022 | 2.977 | 3.000 | 3.022 | 2.978 | 3.000 | 3.022 | 0.955 | – |
| $\omega$ | 0.458 | 0.498 | 0.538 | 0.456 | 0.497 | 0.537 | 0.458 | 0.498 | 0.538 | 0.957 | – |
| $\lambda$ | 2.351 | 3.048 | 4.019 | 2.246 | 2.928 | 3.874 | 2.295 | 2.984 | 3.925 | 0.954 | $4\times10^{-53}$ |
| $\lambda = 4$ | | | | | | | | | | | |
| $\beta_1$ | 4.968 | 5.001 | 5.033 | 4.968 | 5.001 | 5.034 | 4.969 | 5.001 | 5.034 | 0.950 | – |
| $\beta_2$ | 1.981 | 2.000 | 2.017 | 1.981 | 2.000 | 2.019 | 1.981 | 2.000 | 2.018 | 0.961 | – |
| $\beta_3$ | 2.981 | 2.999 | 3.020 | 2.981 | 2.999 | 3.020 | 2.981 | 2.999 | 3.019 | 0.964 | – |
| $\omega$ | 0.460 | 0.499 | 0.536 | 0.458 | 0.498 | 0.537 | 0.459 | 0.499 | 0.536 | 0.941 | – |
| $\lambda$ | 3.142 | 4.073 | 5.415 | 2.996 | 3.895 | 5.119 | 3.073 | 3.976 | 5.216 | 0.954 | $3\times10^{-55}$ |
| $\lambda = 5$ | | | | | | | | | | | |
| $\beta_1$ | 4.974 | 5.000 | 5.027 | 4.974 | 5.000 | 5.029 | 4.974 | 5.000 | 5.028 | 0.962 | – |
| $\beta_2$ | 1.983 | 2.000 | 2.017 | 1.982 | 2.000 | 2.017 | 1.982 | 2.000 | 2.017 | 0.963 | – |
| $\beta_3$ | 2.983 | 2.999 | 3.017 | 2.983 | 2.999 | 3.017 | 2.983 | 2.999 | 3.017 | 0.966 | – |
| $\omega$ | 0.465 | 0.499 | 0.532 | 0.463 | 0.497 | 0.532 | 0.465 | 0.498 | 0.532 | 0.953 | – |
| $\lambda$ | 3.913 | 5.199 | 7.083 | 3.669 | 4.895 | 6.654 | 3.795 | 5.043 | 6.800 | 0.956 | $5\times10^{-46}$ |

Table 3: Point estimators, coverage proportions and Bayes factors: Linear regression model with residual errors simulated from a skew-normal distribution, $n = 500$.

skewness parameter, in both cases, we use the Student-$t$ approximation (with $1/2$ degrees of freedom and scale $\pi/2$) to the reference prior proposed in [8]. For the degrees of freedom we use the prior (10) with hyperparameter $d = 10$. This prior structure allows us to match the priors associated to the different distributional assumptions, which in turn helps us to justify the use of Bayes factors for model comparison. The propriety of the corresponding posterior distributions is guaranteed by Proposition 1 and Corollary 1. We obtain a posterior sample of size $10,000$ after a burn-in period of $25,000$ iterations and thinned to every 25th iteration for each of these models (275,000 iterations in total). Table 4 presents a summary of the posterior samples as well as the BIC, Bayes factors, and LPML. All of the model selection criteria favour the model with skew-$t$ errors, which suggest the presence of heavy tails and skewness.

In order to assess the goodness of fit of the SSt model, we propose a Bayesian residual analysis as follows. First, for each posterior sample of the parameters $(\beta_1^{(k)}, \beta_2^{(k)})$, $k = 1, \ldots, 10000$, we obtain the median of the residual errors $\varepsilon_j^{(k)} = Lbm_j - \beta_1^{(k)} Ht_j - \beta_2^{(k)} Wt_j$. By using the posterior samples of $(\omega, \lambda, \delta)$, we approximate the posterior predictive distribution of the errors and obtain a sample of size $10,000$ from this distribution. Then, we obtain $M = 10,000$ sub-samples of size $102$ from the sample of the predictive distribution, and construct a QQ-plot for each of these samples. Finally, using these $M$ QQ-plots we generate a predictive envelope by taking the 5% and 95% quantiles of the QQ-plots at each sample quantile point. This envelope can be used to visually assess the fit of the residuals and to detect shortcomings of a model in specific regions. Figure 5 shows the predictive quantile envelope associated to both the SSt and normal models. We notice that the envelope associated to the SSt model covers all the data points as well as the straight line, which represents perfect fit. On the other hand, the envelope produced with the normal model does not contain the straight line in several regions.

| Model | SSt | SN | $t$ | N |
|---|---|---|---|---|
| $\beta_1$ | 0.05 (0.03,0.08) | 0.05 (0.04,0.08) | 0.06 (0.04,0.08) | 0.08 (0.06,0.10) |
| $\beta_2$ | 0.81 (0.75,0.87) | 0.81 (0.76,0.85) | 0.77 (0.74,0.81) | 0.73 (0.69,0.77) |
| $\omega$ | 1.85 (1.18,2.77) | 3.51 (2.94,4.22) | 1.24 (0.96,1.62) | 2.29 (2.01,2.64) |
| $\lambda$ | -2.43 (-7.04,-0.57) | -4.08 (-7.89,-2.18) | – | – |
| $\delta$ | 2.69 (1.54,6.03) | – | 2.33 (1.44,4.28) | – |
| BIC | **231.07** | 238.21 | 233.00 | 241.83 |
| Bayes factor | – | $4 \times 10^{-4}$ | $8 \times 10^{-3}$ | $3 \times 10^{-7}$ |
| LPML | **-217.39** | -224.48 | -218.84 | -232.18 |

Table 4: AIS Data: Posterior median, 95% credible intervals, BIC and Bayes factors against the SSt model.

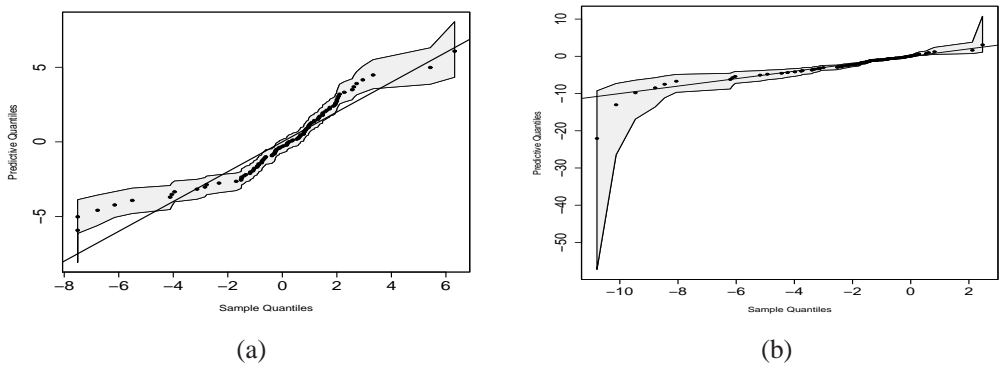

Figure 5: Bayesian residual analysis: (a) skew-normal errors; (b) skew-$t$ errors.

## 5.2 North Central Cancer Treatment Group (NCCTG) data

We now analyse the popular NCCTG lung cancer data. This data set contains the survival times of $n = 227$ patients (the total number of patients is 228 but, for the sake of simplicity, we have removed one patient with a missing covariate) with advanced lung cancer from the NCCTG. The sample contains 63 censored observations. The goal of this study was to compare the descriptive information from a questionnaire applied to a group of patients against the information obtained by each patient's physician, in terms of prognostic power [27]. We fit an AFT model (8) with three covariates "age" (in years), "sex" (Male=1, Female=2), "ph.ecog" (ECOG performance score, 0=good–5=dead) as well as an intercept, together with 4 distributional assumptions on the residual errors. We compare the inference obtained with skew-logistic errors (SS Logistic), skew-normal errors (SS Normal), and the corresponding symmetric sub-models (logistic and normal). We adopt the prior structure (4) with $a = 1$ and the reference prior for $\lambda$ proposed in [32] (which is approximated using a Student-$t$ distribution with 1/2 degrees of freedom and scale 3/4). The propriety of the corresponding posterior distributions is guaranteed by Theorem 2 in the Supporting Web Material. We obtain a posterior sample of size $10,000$ after a burn-in of $50,000$ and thinned to every 25th iteration (300,000 iterations in total). Table 5 shows a summary of the posterior samples as well as the model comparison. The Bayes factors, BIC, and LPML favour the model with SS logistic errors, which suggests the presence of skewness and slightly heavier tails than normal. In this case, we cannot obtain a residual analysis given that the data set contains censored observations. Although there have been some attempts to produce Bayesian residual analyses in the presence of censored observations [12], there does not seem to be an agreement on which of these methods is more appropriate.

| Model | SS logistic | SS Normal | Logistic | Normal |
|---|---|---|---|---|
| Intercept | 6.79 (5.82, 7.74) | 7.19 (6.20, 8.28) | 5.96 ( 4.98, 6.96) | 6.48 (5.30, 7.62) |
| Age | -0.01 (-0.02,0.00) | -0.01 (-0.02,0.00) | -0.01 (-0.02, 0.00) | -0.02 (-0.03, 0.00) |
| Sex | 0.44 (0.19,0.72) | 0.45 (0.17,0.72) | 0.49 (0.22, 0.76) | 0.53 (0.23, 0.84) |
| ph.ecog | -0.37 (-0.55,-0.19) | -0.34 (-0.53, -0.15) | -0.41 (-0.60, -0.22) | -0.36 (-0.57,-0.16) |
| $\omega$ | 0.71 (0.57, 0.86) | 1.46 (1.28, 1.67) | 0.55 (0.48,0.63) | 1.04 (0.93, 1.17) |
| $\lambda$ | -2.03 (-4.20,-0.64) | -3.26 (-5.90, -1.85) | – | – |
| BIC | **555.93** | 563.67 | 562.17 | 580.96 |
| Bayes factor | – | 0.008 | 0.108 | $1 \times 10^{-5}$ |
| LPML | **-268.53** | -272.48 | -272.91 | -283.23 |

Table 5: NCCTG Lung Cancer data: Posterior median, 95% credible intervals, BIC and Bayes factors against the SS logistic model.

Now, using the AFT models with SS Logistic and Logistic errors, we study the remaining life of five patients which survived beyond the end of the study. A proper prediction of the individual remaining life is of great importance in medicine since this information can be used for planning health care. Table 6 presents a summary of the quantiles of these distributions for the first 5 censored patients. These quantiles were obtained by using the posterior AFT model centred at the median (averaged over all the posterior samples). We can observe that the quantiles higher than 50% are much larger in the model with Logistic errors than those obtained with the model with SS Logistic errors. This difference seems to be produced by the larger values in the posterior sample of $\omega$ in the model with Logistic errors, which are overestimated by the lack of flexibility of this error distribution.

## 6 Discussion

We proposed a flexible class of LRMs with SSSMN errors that can capture a variety of tail behaviours and skewness. The proposed models represent an extension to LRM with SMN errors. The latter have been widely used to capture heteroscedasticity and the presence of outliers, but they cannot capture departures from symmetry. The use of SSSMN error distributions can capture additional unobserved heterogeneity which has the effect of inducing asymmetry in the residual errors. We introduced a non-informative prior structure that induces proper posteriors under rather mild conditions. We presented propriety results in a unified framework that covers the usual LRMs and AFT models. We provided tangible conditions for checking the propriety of the posterior distribution in cases when the response variables are censored. We have illustrated the need for this sort of extension with simulated and real data. Our simulation studies also indicate that the proposed prior structure induces posterior distributions with appealing frequentist properties. We have emphasised the usefulness of the proposed models in survival analysis, which are relevant in medicine. However, they can also be applied to other contexts such as finance, biology, or industrial applications.

| Quantile | 5% | 25% | 50% | 75% | 95% |
|---|---|---|---|---|---|
| SS Logistic model | | | | | |
| Patient 1 | 1054.1 | 1163.4 | 1576.7 | 2120.3 | 3539.5 |
| Patient 2 | 1045.8 | 1155.6 | 1347.6 | 1693.3 | 2693.3 |
| Patient 3 | 1008.1 | 1200.1 | 1516.2 | 2048.7 | 3436.6 |
| Patient 4 | 896.0 | 1217.3 | 1706.0 | 2478.9 | 4363.4 |
| Patient 5 | 882.6 | 1071.5 | 1374.5 | 1881.8 | 3174.8 |
| Logistic model | | | | | |
| Patient 1 | 1043.9 | 1211.7 | 1549.4 | 2318.6 | 5700.5 |
| Patient 2 | 1052.9 | 1206.9 | 1520.3 | 2241.5 | 5453.9 |
| Patient 3 | 997.8 | 1160.0 | 1485.5 | 2225.7 | 5481.2 |
| Patient 4 | 857.6 | 1034.8 | 1378.8 | 2136.7 | 5387.6 |
| Patient 5 | 869.3 | 1013.5 | 1302.4 | 1956.2 | 4829.0 |

Table 6: NCCTG Lung Cancer data: Quantiles of the predictive residual life distribution for the Median SS Logistic and Logistic models.

We implemented the proposed models using a general-purpose MCMC sampler. This was possible given that we were using distributions with numerically tractable PDF and CDF. In a more general framework, the Metropolis within Gibbs sampling strategy (which takes advantage of the stochastic representation of SMN) proposed in [36] can be extended to models with SSSMN errors by using the stochastic representation of this sort of distributions [39]. This only implies an additional step in the Gibbs sampler from [36]. In those cases when the PDF and CDF of the SSSMN of interest are tractable, it is possible to implement a Metropolis-within-Gibbs sampler with already available R packages such as 'spBayes' [20].

## Acknowledgements

We thank an Associate Editor and two referees for their very helpful comments. FJR gratefully acknowledges research support from EPSRC grant EP/K007521/1. MGG's research is supported by King Abdullah University of Science and Technology (KAUST).

## References

[1] R. Arellano-Valle, L.M. Castro, M.G. Genton, and H.W. Gómez. Bayesian inference for shape mixtures of skewed distributions, with application to regression analysis. *Bayesian Analysis*, 3: 513–540, 2008.

[2] A. Azzalini. A class of distributions which includes the normal ones. *Scandinavian Journal of Statistics*, 12:171–178, 1985.

[3] A. Azzalini and A. Capitanio. Distributions generated by perturbation of symmetry with emphasis on a multivariate skew-t distribution. *Journal of Royal Statistical Society, Series B*, 65:367–389, 2003.

[4] A. Azzalini and A. Capitanio. *The Skew-Normal and Related Families*. Cambridge University Press, 2014.

[5] A. Azzalini and M.G. Genton. Robust likelihood methods based on the skew-$t$ and related distributions. *International Statistical Review*, 76:106–129, 2008.

[6] A. Azzalini, T. Dal Cappello, and S. Kotz. Log-skew-normal and log-skew-t distributions as models for family income data. *Journal of Income Distribution*, 11:12–20, 2003.

[7] M. Barros, G.A. Paula, and V. Leiva. A new class of survival regression models with heavy-tailed errors: robustness and diagnostics. *Lifetime Data Analysis*, 14:316–332, 2008.

[8] C.L. Bayes and M.D. Branco. Bayesian inference for the skewness parameter of the scalar skew-normal distribution. *Brazilian Journal of Probability and Statistics*, 21:141–163, 2007.

[9] M.D. Branco, M.G. Genton, and B. Liseo. Objective Bayesian analysis of skew-t distributions. *Scandinavian Journal of Statistics*, 40:63–85, 2012.

[10] S. Cabras, W. Racugno, M.E. Castellanos, and L. Ventura. A matching prior for the shape parameter of the skew-normal distribution. *Scandinavian Journal of Statistics*, 39(2):236–247, 2012.

[11] A. Canale and B. Scarpa. Informative Bayesian inference for the skew-normal distribution. *arXiv preprint arXiv:1305.3080*, 2013.

[12] K. Chaloner. Bayesian residual analysis in the presence of censoring. *Biometrika*, 78:637–644, 1991.

[13] H.M. Choi and J.P. Hobert. Analysis of MCMC algorithms for Bayesian linear regression with Laplace errors. *Journal of Multivariate Analysis*, 117:32–40, 2013.

[14] J.A. Christen and C. Fox. A general purpose sampling algorithm for continuous distributions (the twalk). *Bayesian Analysis*, 5:263–282, 2010.

[15] J.M. Dickey. The weighted likelihood ratio, linear hypotheses on normal location parameters. *The Annals of Mathematical Statistics*, pages 204–223, 1971.

[16] C. Fernández and M.F.J. Steel. Multivariate Student-$t$ regression models: Pitfalls and inference. *Biometrika*, 86:153–167, 1999.

[17] C. Fernández and M.F.J. Steel. On the dangers of modelling through continuous distributions: A Bayesian perspective. In J. M. Bernardo, J. O. Berger, A. P. Dawid, and A. F. M. Smith, editors, *Bayesian Statistics 6*. Oxford University Press, Oxford, 1999.

[18] C. Fernández and M.F.J. Steel. Bayesian regression analysis with scale mixtures of normals. *Econometric Theory*, 80:80–101, 2000.

[19] J.T.A.S. Ferreira and M.F.J. Steel. A new class of skewed multivariate distributions with applications to regression analysis. *Statistica Sinica*, 17:505–529, 2007.

[20] A.O. Finley, S. Banerjee, and B.P. Carlin. spbayes: An R package for univariate and multivariate hierarchical point-referenced spatial models. *Journal of Statistical Software*, 19:1–24, 2007.

[21] T.C.O. Fonseca, M.A.R. Ferreira, and H.S. Migon. Objective Bayesian analysis for the Student-$t$ regression model. *Biometrika*, 95:325–333, 2008.

[22] M.G. Genton. *Skew-Elliptical Distributions and Their Applications: A Journey Beyond Normality*. Chapman & Hall/CRC, 2004.

[23] M.A. Juárez and M.F.J. Steel. Model-based clustering of non-Gaussian panel data based on skew-$t$ distributions. *Journal of Business and Economic Statistics*, 28:52–66, 2010.

[24] Y.J. Jung and J.P. Hobert. Spectral properties of MCMC algorithms for Bayesian linear regression with generalized hyperbolic errors. *Statistics & Probability Letters*, 95:92–100, 2014.

[25] G.D. Lin and J. Stoyanov. The logarithmic skew-normal distributions are moment–indeterminate. *Journal of Applied Probability*, 46:909–916, 2009.

[26] B. Liseo and N. Loperfido. A note on reference priors for the scalar skew-normal distribution. *Journal of Statistical Planning and Inference*, 136:373–389, 2006.

[27] C.L. Loprinzi, J.A. Laurie, H.S. Wieand, J.E. Krook, P.J. Novotny, J.W. Kugler, J. Bartel, M. Law, M. Bateman, N.E. Klatt, and et al. Prospective evaluation of prognostic variables from patient-completed questionnaires. *Journal of Clinical Oncology*, 23:601–607, 1994.

[28] Y.V. Marchenko and M.G. Genton. Multivariate log-skew-elliptical distributions with applications to precipitation data. *Environmetrics*, 21:318–340, 2010.

[29] R Core Team. *R: A Language and Environment for Statistical Computing*. R Foundation for Statistical Computing, Vienna, Austria, 2015. URL http://www.R-project.org/.

[30] V. Roy and D.K. Dey. Propriety of posterior distributions arising in categorical and survival models under generalized extreme value distribution. *Statistica Sinica*, 24:699–722, 2014.

[31] F.J. Rubio and Y. Hong. Survival and lifetime data analysis with a flexible class of distributions. *Journal of Applied Statistics*, in press.

[32] F.J. Rubio and B. Liseo. On the independence Jeffreys prior for skew-symmetric models. *Statistics & Probability Letters*, 85:91–97, 2014.

[33] F.J. Rubio and M.F.J. Steel. Bayesian modelling of skewness and kurtosis with two-piece scale and shape distributions. *Electronic Journal of Statistics*, 9(2):1884–1912, 2015.

[34] S.J. Sahu, D.K. Dey, and M.D. Branco. A new class of multivariate skew distributions with applications to Bayesian regression models. *Canadian Journal of Statistics*, 31:129–150, 2003.

[35] D. Sinha, M.-H. Chen, and S.K. Ghosh. Bayesian analysis and model selection for interval-censored survival data. *Biometrics*, pages 585–590, 1999.

[36] C. Vallejos and M.F.J. Steel. Objective Bayesian survival analysis using shape mixtures of log-normal distributions. *Journal of the American Statistical Association*, 110:697–710, 2015.

[37] A.S. Wahed and M.M. Ali. The skew logistic distribution. *Journal of Statistical Research*, 35: 71–80, 2001.

[38] J. Wang and M.G. Genton. The multivariate skew-slash distribution. *Journal of Statistical Planning and Inference*, 136:209–220, 2006.

[39] J. Wang, J. Boyer, and M.G. Genton. A skew symmetric representation of multivariate distributions. *Statistica Sinica*, 14:1259–1270, 2004.

# Supporting Web Material: Propriety results and additional simulation study.

Francisco J. Rubio[a*] and Marc G. Genton[b]

[a] University of Warwick, Department of Statistics, Coventry, CV4 7AL, UK.

[b] CEMSE Division, King Abdullah University of Science and Technology, Thuwal 23955-6900, Saudi Arabia.

* E-mail: Francisco.Rubio@warwick.ac.uk

## 1 Propriety results

In the material below, equation numbers refer to equations in the main paper. Equations associated to the Supporting Web Material have an "S" suffixed to the equation number.

**Proposition 1** *Consider the model (3)–(4). Assume that $f$ in (2) is a scale mixture of normals and that $\mathrm{rank}(\mathbf{X}) = p$. Then, the posterior of $(\boldsymbol{\beta}, \omega, \lambda, \delta)$ is proper if the posterior distribution associated to model (3), with errors distributed according to a scale mixture of normals $f$ with shape parameter $\delta$, scale parameter $\omega$, and prior structure given by:*

$$\pi(\boldsymbol{\beta}, \omega, \delta) \propto \frac{\tilde{p}(\delta)}{\omega^a}, \tag{1S}$$

*where $\tilde{p}(\delta)$, the marginal prior on $\delta$, is proper.*

**Corollary 1** *Consider the model (3)–(4) and assume that $f$ in (2) is a scale mixture of normals. Then,*

  (i) *If $a = 1$, a sufficient condition for the propriety of the posterior of $(\boldsymbol{\beta}, \omega, \delta, \lambda)$ is $n > p$.*

  (ii) *If $a > 1$, a sufficient condition for the propriety of the posterior is $n > p + 1 - a$ and*

$$\int \tau^{-\frac{a-1}{2}} \tilde{p}(\delta) dH(\tau|\delta) d\delta < \infty,$$

  *where $\tilde{p}(\delta)$ represents the marginal prior on $\delta$.*

**Theorem 1** *Suppose that $T_j | \mathbf{x}_j, \boldsymbol{\beta}, \omega, \lambda, \delta \sim \mathrm{LSS}(\mathbf{x}_j^\top \boldsymbol{\beta}, \omega, \lambda, \delta; f, g)$, that $f$ in (9) is a scale mixture of normals, and that $\mathrm{rank}(\mathbf{X}) = p$. Consider the prior structure (4) for this model. Suppose also that $n_c \leq n$ observations are censored and $n_o = n - n_c$ are observed. Then,*

  (i) *If $a = 1$, a sufficient condition for the propriety of the corresponding posterior is $n_o > p$.*

  (ii) *If $a > 1$, a sufficient condition for the propriety of the posterior is $n_o > p + 1 - a$ and (5).*

An extreme case arises when all of the observations are censored. The following result provides conditions for the propriety of the posterior under this scenario.

**Theorem 2** *Consider the model (8) with prior (4), where $f$ is a scale mixture of normals. Suppose that $n_I \leq n$ observations are interval censored, where the length of these intervals is finite, and that the other $n - n_I$ observations are censored (not necessarily in a finite interval). Denote the $n_I$ interval-censored observations as $(I_1, \ldots, I_{n_I})$, and let $\mathbf{X}_{n_I}$ be the corresponding design submatrix. Then, the corresponding posterior is proper if $\mathcal{E} = I_1 \times \cdots \times I_{n_I}$ and the column space of $\mathbf{X}_{n_I}$ are disjoint, together with one of the following conditions*

(a) $a = 1$ and $n_I > p$.

(b) $a > 1$, $n_I > p + 1 - a$, and (5).

Theorem 2 represents a generalisation of Theorem 5 from [6] and also provides more tangible conditions on the type of censoring required to guarantee the existence of the posterior distribution. An alternative way of checking that $\mathcal{E}$ and the column space of $\mathbf{X}_{n_I}$ are disjoint consists of formulating this condition as a linear programming (LP) problem. Denote $\boldsymbol{\eta} \in \mathbb{R}^p$, $\boldsymbol{\psi} = (\psi_1, \ldots, \psi_{n_I})^\top \in \mathcal{E}$ and $I_j = [l_j, u_j]$, $j = 1, \ldots, n_I$. Define the LP problem:

$$\begin{aligned}
&\text{Find} && \max_{\boldsymbol{\eta}, \boldsymbol{\psi}} \; 1, \\
&\text{Subject to} && \mathbf{X}_{n_I} \boldsymbol{\eta} = \boldsymbol{\psi}, \\
&\text{and} && \log l_j \leq \psi_j \leq \log u_j, \; j = 1, \ldots, n_I.
\end{aligned} \quad (2S)$$

Thus, condition (iii) is equivalent to verifying the infeasibility of the LP problem (2S), for which there are several theoretical and numerical tools (LP solvers) available (see *e.g.* [3]). Note also that the aim of this formulation is not to solve the maximisation problem (which is trivial), but to check the feasibility of the restrictions. This is if the LP solver reports infeasibility, then the sets are disjoint, otherwise it should return some feasible vector.

## 2 Extensions to multivariate linear regression

In this section, we present a multivariate extension of the Bayesian LRM discussed in Section 2. Recall first that a random vector $\mathbf{Z}$ is said to be distributed according to a $m-$variate skew-symmetric distribution, $m \geq 1$, if its PDF can be written as [7]:

$$s(\mathbf{z}|\boldsymbol{\xi}) = 2f(\mathbf{z} - \boldsymbol{\xi})\varphi(\mathbf{z} - \boldsymbol{\xi}), \; \mathbf{z} \in \mathbb{R}^m, \quad (3S)$$

where $f$ is a symmetric PDF with support on $\mathbb{R}^m$, and $\varphi : \mathbb{R}^m \to [0, 1]$ satisfies the condition $\varphi(\mathbf{z}) = 1 - \varphi(-\mathbf{z})$. This class of distributions contains the multivariate skew-normal [2] and the multivariate skew-$t$ distributions [1] as particular cases.

Consider now the linear regression model:

$$\mathbf{y}_j = \mathbf{B}^\top \mathbf{x}_j + \boldsymbol{\varepsilon}_j, \; j = 1, \ldots, n, \quad (4S)$$

where $\mathbf{y}_j \in \mathbb{R}^m$, $\mathbf{B}$ is a $p \times m$ matrix of regression parameters, and $\mathbf{x}_j$ is a known $p \times 1$ vector of covariates. Let $\mathbf{X} = (\mathbf{x}_1, \ldots, \mathbf{x}_n)^\top$ denote the entire design matrix, and suppose that this matrix has full column rank. We focus on the study of the model (4S) with errors distributed according to a certain class of skew-symmetric distributions. In order to introduce this model, recall that a PDF $f$ is said to be a multivariate scale mixture of normals (MSMN), if it can be written as follows:

$$f(\mathbf{z}|\boldsymbol{\xi}, \boldsymbol{\Sigma}, \delta) = \int_0^\infty \frac{\tau^{m/2}}{(2\pi)^{m/2}|\boldsymbol{\Sigma}|^{1/2}} \exp\left\{-\frac{\tau}{2}(\mathbf{z} - \boldsymbol{\xi})^\top \boldsymbol{\Sigma}^{-1} (\mathbf{z} - \boldsymbol{\xi})\right\} dH(\tau|\delta), \; \mathbf{z} \in \mathbb{R}^m,$$

where $H$ is a mixing distribution, $\delta \in \Delta \subset \mathbb{R}$ is a shape parameter, and $\boldsymbol{\Sigma}$ is a $m \times m$ positive definite symmetric matrix. This family contains the multivariate normal and the multivariate Student-$t$ distributions, among others. We say that a density $s$ is a multivariate skew-symmetric scale mixture of normals (MSSSMN) if it can be written as in (3S) where $f$ is a MSMN.

**Theorem 3** *Consider the linear regression model (4S) and suppose that the errors*
$\boldsymbol{\varepsilon}_j \overset{i.i.d.}{\sim} \text{MSSSMN}(\mathbf{0}, \boldsymbol{\Sigma}, \boldsymbol{\lambda}, \delta; f, \varphi)$, *where $f$ is a MSMN with shape parameter $\delta$, and the skewing function $\varphi$ contains a skewness parameter vector $\boldsymbol{\lambda}$. Adopt the prior structure:*

$$\pi(\boldsymbol{\beta}, \boldsymbol{\Sigma}, \boldsymbol{\lambda}, \delta) \propto \frac{p(\boldsymbol{\lambda}, \delta)}{\det(\boldsymbol{\Sigma})^{\frac{m+1}{2}}}, \quad (5S)$$

*where $p(\boldsymbol{\lambda}, \delta)$ is assumed to be proper. Then, the posterior distribution is proper, for almost any sample, provided that $n \geq m + p$ and $\text{rank}(\mathbf{X}) = p$.*

This result indicates that the introduction of skewness, via the skew-symmetric construction (3S), does not affect the existence of the posterior, provided that the prior on the skewness parameter vector $\boldsymbol{\lambda}$ is proper. The prior structure (5S) is of interest in practice given that it is a generalisation of the structure of the independence Jeffreys prior (see [4] for the structure of this prior in the symmetric case). Moreover, it provides a general structure that leads to proper posteriors under rather mild conditions. Analogously to the results presented for univariate responses, this result holds with probability one. We refer the reader to [4] for a discussion on some zero-probability events that produce improper posteriors. The following results present particular details on the propriety of the posterior for skew-normal and skew-$t$ errors

**Corollary 2** *Consider the linear regression model (4S) and suppose that the errors are distributed according to the multivariate skew-normal distribution [2]. Adopt the prior structure:*

$$\pi(\boldsymbol{\beta}, \boldsymbol{\Sigma}, \boldsymbol{\lambda}) \propto \frac{p(\boldsymbol{\lambda})}{\det(\boldsymbol{\Sigma})^{\frac{m+1}{2}}},$$

*where $p(\boldsymbol{\lambda})$ is assumed to be proper. Then, the posterior distribution is proper provided that $n \geq m + p$ and $\text{rank}(\mathbf{X}) = p$.*

**Corollary 3** *Consider the linear regression model (4S) and suppose that the errors are distributed according to the multivariate skew-$t$ distribution [1] with unknown degrees of freedom $\delta > 0$. Adopt the prior structure:*

$$\pi(\boldsymbol{\beta}, \boldsymbol{\Sigma}, \boldsymbol{\lambda}, \delta) \propto \frac{p(\boldsymbol{\lambda})p(\delta)}{\det(\boldsymbol{\Sigma})^{\frac{m+1}{2}}},$$

*where $p(\boldsymbol{\lambda})$ and $p(\delta)$ are assumed to be proper. Then, the posterior distribution is proper provided that $n \geq m + p$ and $\text{rank}(\mathbf{X}) = p$.*

## Proofs

### Proof of Proposition 1

The marginal likelihood of the data is given by

$$m(\mathbf{y}) = \int \left[\prod_{j=1}^{n} s(y_j | \mathbf{x}_j^\top \boldsymbol{\beta}, \omega, \delta, \lambda)\right] \frac{p(\lambda, \delta)}{\omega^a} d\boldsymbol{\beta} d\omega d\delta d\lambda.$$

Using that $0 \leq G(\cdot) \leq 1$ in (2) and that $p(\lambda, \delta)$ is proper it follows that

$$\begin{aligned} m(\mathbf{y}) &\leq \int \left[\prod_{j=1}^{n} \frac{2}{\omega} f\left(\left.\frac{y_j - \mathbf{x}_j^\top \boldsymbol{\beta}}{\omega}\right|\delta\right)\right] \frac{p(\lambda, \delta)}{\omega^a} d\boldsymbol{\beta} d\omega d\delta d\lambda \\ &= 2^n \int \left[\prod_{j=1}^{n} \frac{1}{\omega} f\left(\left.\frac{y_j - \mathbf{x}_j^\top \boldsymbol{\beta}}{\omega}\right|\delta\right)\right] \frac{\tilde{p}(\delta)}{\omega^a} d\boldsymbol{\beta} d\omega d\delta, \end{aligned} \quad (6S)$$

where $\tilde{p}(\delta) = \int p(\lambda, \delta) d\lambda$, which is proper. The result follows by noting that the last term in inequality (6S) corresponds to the marginal likelihood of a LRM with residual errors distributed according to the symmetric distribution $f$ and prior structure $\pi(\boldsymbol{\beta}, \omega, \delta) \propto \tilde{p}(\delta)/\omega^a$. □

**Proof of Corollary 1**

(i) follows by Proposition 1 together with Theorem 1 in [5]. (ii) follows by Proposition 1 together with the proofs of Lemma 1 from [4] and Theorem 2 from [6]. □

**Proof of Theorem 1**

As discussed in [6], the contribution of a censored observation to the likelihood function is a factor in $[0,1]$. Then, the marginal likelihood of the complete sample is upper bounded by the marginal likelihood of the uncensored observations. Therefore, the propriety of the posterior distribution can be based on the uncensored observations. □

**Proof of Theorem 2**

Provided that either (i) or (ii) are satisfied, it follows that

$$I(t_1,\ldots,t_{n_I}) = \int \prod_{j=1}^{n_I} s_l(t_j|\mathbf{x}_j^\top \boldsymbol{\beta}, \omega, \lambda, \delta) \pi(\boldsymbol{\beta}, \omega, \lambda, \delta) d\boldsymbol{\beta} d\omega d\lambda d\delta < \infty,$$

for each $(t_1,\ldots,t_{n_I}) \in \mathcal{E}$. Finally, given that the Lebesgue measure of $\mathcal{E}$ is finite (since this set is bounded), it follows that the marginal likelihood of the data $\int_{\mathcal{E}} I(t_1,\ldots,t_{n_I}) dt_1 \cdots dt_{n_I}$ is finite. □

**Proof of Theorem 3**

The proof follows analogously to the proof of Proposition 1 by noting that the inequality $2f(\mathbf{x})\varphi(\mathbf{x}) \leq 2f(\mathbf{x})$ holds. □

# 3 Simulation with skew-logistic and skew-$t$ errors.

For a description of the simulation scenario see Section 4 of the main paper.

| Method | MLE | | | MAP | | | Median | | | Coverage | BF |
|---|---|---|---|---|---|---|---|---|---|---|---|
| | 5% | 50% | 95% | 5% | 50% | 95% | 5% | 50% | 95% | | |
| $\lambda = 0$ | | | | | | | | | | | |
| $\beta_1$ | 4.377 | 4.971 | 5.607 | 4.364 | 4.966 | 5.596 | 4.463 | 4.966 | 5.512 | 0.960 | – |
| $\beta_2$ | 1.846 | 1.998 | 2.150 | 1.841 | 1.999 | 2.155 | 1.845 | 1.998 | 2.149 | 0.951 | – |
| $\beta_3$ | 2.862 | 2.999 | 3.147 | 2.862 | 3.001 | 3.154 | 2.861 | 2.999 | 3.148 | 0.960 | – |
| $\omega$ | 0.442 | 0.523 | 0.666 | 0.458 | 0.529 | 0.627 | 0.476 | 0.553 | 0.652 | 0.904 | – |
| $\lambda$ | -1.114 | 0.037 | 1.194 | -0.720 | 0.026 | 0.709 | -0.861 | 0.038 | 0.895 | 0.961 | 2.817 |
| $\lambda = 1$ | | | | | | | | | | | |
| $\beta_1$ | 4.684 | 4.993 | 5.484 | 4.695 | 5.020 | 5.530 | 4.745 | 5.065 | 5.506 | 0.959 | – |
| $\beta_2$ | 1.884 | 2.001 | 2.110 | 1.881 | 2.004 | 2.121 | 1.880 | 2.001 | 2.111 | 0.958 | – |
| $\beta_3$ | 2.886 | 3.003 | 3.122 | 2.884 | 3.005 | 3.131 | 2.884 | 3.004 | 3.123 | 0.954 | – |
| $\omega$ | 0.381 | 0.493 | 0.654 | 0.387 | 0.469 | 0.625 | 0.405 | 0.490 | 0.627 | 0.979 | – |
| $\lambda$ | 0.014 | 1.086 | 2.924 | -0.053 | 0.593 | 2.006 | -0.044 | 0.816 | 2.305 | 0.958 | 1.390 |
| $\lambda = 2$ | | | | | | | | | | | |
| $\beta_1$ | 4.812 | 4.996 | 5.317 | 4.807 | 5.011 | 5.346 | 4.831 | 5.050 | 5.359 | 0.942 | – |
| $\beta_2$ | 1.896 | 2.000 | 2.094 | 1.893 | 2.001 | 2.098 | 1.896 | 2.000 | 2.095 | 0.959 | – |
| $\beta_3$ | 2.896 | 2.998 | 3.092 | 2.893 | 2.999 | 3.091 | 2.898 | 2.998 | 3.091 | 0.959 | – |
| $\omega$ | 0.367 | 0.496 | 0.617 | 0.350 | 0.467 | 0.606 | 0.372 | 0.474 | 0.605 | 0.958 | – |
| $\lambda$ | 0.664 | 2.260 | 5.121 | 0.322 | 1.466 | 3.443 | 0.548 | 1.772 | 4.236 | 0.947 | 0.307 |
| $\lambda = 3$ | | | | | | | | | | | |
| $\beta_1$ | 4.859 | 5.002 | 5.230 | 4.854 | 5.008 | 5.246 | 4.872 | 5.030 | 5.277 | 0.920 | – |
| $\beta_2$ | 1.915 | 2.002 | 2.087 | 1.907 | 2.001 | 2.088 | 1.914 | 2.000 | 2.088 | 0.955 | – |
| $\beta_4$ | 2.913 | 3.000 | 3.087 | 2.910 | 3.000 | 3.086 | 2.915 | 3.000 | 3.086 | 0.968 | – |
| $\omega$ | 0.374 | 0.492 | 0.601 | 0.342 | 0.476 | 0.593 | 0.361 | 0.478 | 0.590 | 0.933 | – |
| $\lambda$ | 1.213 | 3.225 | 9.416 | 0.580 | 2.305 | 5.027 | 0.912 | 2.660 | 6.595 | 0.923 | 0.078 |
| $\lambda = 4$ | | | | | | | | | | | |
| $\beta_1$ | 4.874 | 4.999 | 5.156 | 4.867 | 5.004 | 5.182 | 4.880 | 5.017 | 5.206 | 0.927 | – |
| $\beta_2$ | 1.925 | 1.997 | 2.081 | 1.923 | 1.998 | 2.080 | 1.926 | 1.997 | 2.077 | 0.959 | – |
| $\beta_3$ | 2.921 | 2.996 | 3.075 | 2.919 | 2.996 | 3.081 | 2.924 | 2.997 | 3.075 | 0.961 | – |
| $\omega$ | 0.398 | 0.498 | 0.603 | 0.369 | 0.488 | 0.599 | 0.385 | 0.489 | 0.600 | 0.938 | – |
| $\lambda$ | 1.991 | 4.457 | 107.873 | -0.893 | 3.108 | 7.313 | 1.505 | 3.748 | 13.127 | 0.918 | 0.020 |
| $\lambda = 5$ | | | | | | | | | | | |
| $\beta_1$ | 4.892 | 4.999 | 5.138 | 4.885 | 4.998 | 5.156 | 4.899 | 5.006 | 5.180 | 0.929 | – |
| $\beta_2$ | 1.931 | 2.002 | 2.078 | 1.927 | 2.001 | 2.075 | 1.930 | 2.001 | 2.075 | 0.957 | – |
| $\beta_3$ | 2.924 | 3.001 | 3.074 | 2.925 | 3.001 | 3.076 | 2.926 | 3.001 | 3.070 | 0.958 | – |
| $\omega$ | 0.400 | 0.499 | 0.591 | 0.384 | 0.489 | 0.588 | 0.390 | 0.495 | 0.590 | 0.937 | – |
| $\lambda$ | 2.574 | 5.741 | 144.920 | -3.665 | 3.788 | 12.203 | 2.031 | 4.879 | 18.220 | 0.919 | 0.015 |

Table 1S: Point estimators, coverage proportions and Bayes factors: Linear regression model with residual errors simulated from a skew-logistic distribution, $n = 100$.

| Method | MLE | | | MAP | | | Median | | | Coverage | BF |
|---|---|---|---|---|---|---|---|---|---|---|---|
| | 5% | 50% | 95% | 5% | 50% | 95% | 5% | 50% | 95% | | |
| $\lambda = 0$ | | | | | | | | | | | |
| $\beta_1$ | 4.614 | 5.004 | 5.417 | 4.572 | 5.010 | 5.435 | 4.616 | 5.016 | 5.395 | 0.941 | – |
| $\beta_2$ | 1.911 | 2.000 | 2.089 | 1.903 | 2.001 | 2.092 | 1.910 | 2.001 | 2.088 | 0.959 | – |
| $\beta_3$ | 2.914 | 3.001 | 3.098 | 2.911 | 3.003 | 3.100 | 2.914 | 3.002 | 3.100 | 0.956 | – |
| $\omega$ | 0.464 | 0.510 | 0.587 | 0.470 | 0.517 | 0.575 | 0.480 | 0.528 | 0.589 | 0.917 | – |
| $\lambda$ | -0.600 | -0.020 | 0.603 | -0.538 | -0.020 | 0.488 | -0.595 | -0.023 | 0.573 | 0.935 | 4.669 |
| $\lambda = 1$ | | | | | | | | | | | |
| $\beta_1$ | 4.784 | 4.999 | 5.306 | 4.786 | 5.013 | 5.312 | 4.820 | 5.043 | 5.294 | 0.969 | – |
| $\beta_2$ | 1.922 | 2.000 | 2.070 | 1.920 | 2.000 | 2.076 | 1.922 | 2.000 | 2.070 | 0.957 | – |
| $\beta_3$ | 2.929 | 3.001 | 3.078 | 2.926 | 2.999 | 3.078 | 2.928 | 3.000 | 3.078 | 0.948 | – |
| $\omega$ | 0.409 | 0.499 | 0.598 | 0.404 | 0.474 | 0.586 | 0.419 | 0.487 | 0.585 | 0.967 | – |
| $\lambda$ | 0.280 | 1.028 | 1.942 | 0.217 | 0.785 | 1.718 | 0.314 | 0.885 | 1.738 | 0.965 | 0.405 |
| $\lambda = 2$ | | | | | | | | | | | |
| $\beta_1$ | 4.879 | 5.011 | 5.182 | 4.882 | 5.014 | 5.187 | 4.888 | 5.034 | 5.218 | 0.925 | – |
| $\beta_2$ | 1.941 | 2.001 | 2.059 | 1.938 | 2.000 | 2.062 | 1.940 | 2.001 | 2.059 | 0.962 | – |
| $\beta_3$ | 2.941 | 2.999 | 3.058 | 2.940 | 3.001 | 3.058 | 2.940 | 2.999 | 3.058 | 0.961 | – |
| $\omega$ | 0.413 | 0.492 | 0.569 | 0.389 | 0.483 | 0.568 | 0.402 | 0.483 | 0.564 | 0.941 | – |
| $\lambda$ | 1.060 | 1.993 | 3.219 | 0.777 | 1.743 | 2.861 | 0.910 | 1.813 | 2.999 | 0.922 | 0.003 |
| $\lambda = 3$ | | | | | | | | | | | |
| $\beta_1$ | 4.914 | 5.003 | 5.106 | 4.913 | 5.004 | 5.117 | 4.921 | 5.012 | 5.129 | 0.946 | – |
| $\beta_2$ | 1.949 | 2.001 | 2.051 | 1.948 | 2.000 | 2.053 | 1.950 | 2.001 | 2.051 | 0.963 | – |
| $\beta_3$ | 2.947 | 2.999 | 3.051 | 2.947 | 2.999 | 3.056 | 2.948 | 2.999 | 3.052 | 0.959 | – |
| $\omega$ | 0.430 | 0.496 | 0.565 | 0.422 | 0.492 | 0.562 | 0.421 | 0.493 | 0.561 | 0.936 | – |
| $\lambda$ | 1.942 | 3.136 | 4.910 | 1.676 | 2.801 | 4.328 | 1.752 | 2.914 | 4.584 | 0.937 | $1 \times 10^{-8}$ |
| $\lambda = 4$ | | | | | | | | | | | |
| $\beta_1$ | 4.923 | 4.999 | 5.077 | 4.923 | 5.001 | 5.080 | 4.929 | 5.004 | 5.085 | 0.952 | – |
| $\beta_2$ | 1.954 | 1.999 | 2.046 | 1.953 | 1.998 | 2.049 | 1.955 | 1.998 | 2.047 | 0.955 | – |
| $\beta_3$ | 2.952 | 2.999 | 3.046 | 2.952 | 2.999 | 3.046 | 2.952 | 3.000 | 3.046 | 0.967 | – |
| $\omega$ | 0.439 | 0.499 | 0.560 | 0.432 | 0.495 | 0.556 | 0.434 | 0.496 | 0.559 | 0.948 | – |
| $\lambda$ | 2.788 | 4.161 | 6.653 | 2.487 | 3.717 | 5.715 | 2.580 | 3.924 | 6.203 | 0.953 | $2 \times 10^{-11}$ |
| $\lambda = 5$ | | | | | | | | | | | |
| $\beta_1$ | 4.940 | 4.999 | 5.070 | 4.937 | 5.000 | 5.073 | 4.942 | 5.002 | 5.075 | 0.957 | – |
| $\beta_2$ | 1.958 | 2.000 | 2.043 | 1.956 | 2.001 | 2.042 | 1.958 | 2.000 | 2.043 | 0.965 | – |
| $\beta_3$ | 2.958 | 2.999 | 3.041 | 2.958 | 2.998 | 3.042 | 2.959 | 2.998 | 3.041 | 0.971 | – |
| $\omega$ | 0.442 | 0.499 | 0.556 | 0.439 | 0.496 | 0.555 | 0.441 | 0.497 | 0.555 | 0.958 | – |
| $\lambda$ | 3.432 | 5.271 | 8.539 | 3.002 | 4.631 | 7.160 | 3.233 | 4.919 | 7.875 | 0.950 | $2 \times 10^{-11}$ |

Table 2S: Point estimators, coverage proportions and Bayes factors: Linear regression model with residual errors simulated from a skew-logistic distribution, $n = 250$.

| Method | MLE | | | MAP | | | Median | | | Coverage | BF |
|---|---|---|---|---|---|---|---|---|---|---|---|
| | 5% | 50% | 95% | 5% | 50% | 95% | 5% | 50% | 95% | | |
| $\lambda = 0$ | | | | | | | | | | | |
| $\beta_1$ | 4.730 | 4.993 | 5.254 | 4.721 | 4.995 | 5.272 | 4.715 | 4.991 | 5.270 | 0.926 | – |
| $\beta_2$ | 1.932 | 1.999 | 2.065 | 1.933 | 1.999 | 2.067 | 1.933 | 1.999 | 2.065 | 0.950 | – |
| $\beta_3$ | 2.935 | 3.002 | 3.064 | 2.935 | 3.002 | 3.068 | 2.935 | 3.002 | 3.064 | 0.956 | – |
| $\omega$ | 0.473 | 0.504 | 0.543 | 0.478 | 0.509 | 0.545 | 0.482 | 0.514 | 0.551 | 0.947 | – |
| $\lambda$ | -0.331 | 0.006 | 0.369 | -0.328 | 0.011 | 0.328 | -0.369 | 0.008 | 0.396 | 0.928 | 7.415 |
| $\lambda = 1$ | | | | | | | | | | | |
| $\beta_1$ | 4.832 | 5.006 | 5.235 | 4.828 | 5.009 | 5.226 | 4.851 | 5.034 | 5.226 | 0.946 | – |
| $\beta_2$ | 1.948 | 2.000 | 2.054 | 1.945 | 1.999 | 2.053 | 1.947 | 2.000 | 2.055 | 0.952 | – |
| $\beta_3$ | 2.946 | 3.001 | 3.054 | 2.945 | 2.999 | 3.057 | 2.947 | 3.000 | 3.055 | 0.943 | – |
| $\omega$ | 0.429 | 0.499 | 0.572 | 0.422 | 0.488 | 0.568 | 0.434 | 0.492 | 0.565 | 0.948 | – |
| $\lambda$ | 0.431 | 0.995 | 1.677 | 0.364 | 0.887 | 1.578 | 0.460 | 0.908 | 1.580 | 0.946 | 0.03 |
| $\lambda = 2$ | | | | | | | | | | | |
| $\beta_1$ | 4.918 | 5.003 | 5.111 | 4.919 | 5.005 | 5.121 | 4.924 | 5.013 | 5.132 | 0.947 | – |
| $\beta_2$ | 1.954 | 2.000 | 2.044 | 1.952 | 2.000 | 2.044 | 1.954 | 1.999 | 2.044 | 0.943 | – |
| $\beta_3$ | 2.957 | 2.999 | 3.042 | 2.956 | 2.999 | 3.043 | 2.957 | 2.998 | 3.042 | 0.958 | – |
| $\omega$ | 0.444 | 0.498 | 0.546 | 0.435 | 0.493 | 0.545 | 0.438 | 0.492 | 0.543 | 0.955 | – |
| $\lambda$ | 1.355 | 2.019 | 2.740 | 1.183 | 1.894 | 2.628 | 1.239 | 1.921 | 2.653 | 0.948 | $2\times10^{-16}$ |
| $\lambda = 3$ | | | | | | | | | | | |
| $\beta_1$ | 4.938 | 5.002 | 5.068 | 4.939 | 5.004 | 5.072 | 4.939 | 5.006 | 5.077 | 0.950 | – |
| $\beta_2$ | 1.964 | 1.999 | 2.036 | 1.960 | 1.999 | 2.036 | 1.963 | 2.000 | 2.035 | 0.961 | – |
| $\beta_3$ | 2.961 | 2.999 | 3.037 | 2.961 | 2.999 | 3.039 | 2.962 | 2.998 | 3.036 | 0.953 | – |
| $\omega$ | 0.452 | 0.498 | 0.542 | 0.451 | 0.495 | 0.543 | 0.450 | 0.496 | 0.543 | 0.956 | – |
| $\lambda$ | 2.270 | 2.990 | 4.101 | 2.125 | 2.867 | 3.857 | 2.161 | 2.910 | 3.960 | 0.950 | $4\times10^{-40}$ |
| $\lambda = 4$ | | | | | | | | | | | |
| $\beta_1$ | 4.948 | 5.002 | 5.056 | 4.948 | 5.003 | 5.059 | 4.949 | 5.004 | 5.059 | 0.950 | – |
| $\beta_2$ | 1.966 | 2.000 | 2.034 | 1.966 | 2.000 | 2.034 | 1.966 | 2.000 | 2.033 | 0.952 | – |
| $\beta_3$ | 2.968 | 3.001 | 3.034 | 2.967 | 3.001 | 3.035 | 2.967 | 3.000 | 3.034 | 0.949 | – |
| $\omega$ | 0.457 | 0.499 | 0.542 | 0.454 | 0.498 | 0.543 | 0.455 | 0.498 | 0.543 | 0.955 | – |
| $\lambda$ | 3.062 | 4.081 | 5.598 | 2.884 | 3.846 | 5.311 | 2.968 | 3.961 | 5.361 | 0.946 | $7\times10^{-43}$ |
| $\lambda = 5$ | | | | | | | | | | | |
| $\beta_1$ | 4.958 | 4.999 | 5.051 | 4.957 | 5.000 | 5.051 | 4.958 | 5.001 | 5.053 | 0.958 | – |
| $\beta_2$ | 1.968 | 1.999 | 2.028 | 1.967 | 1.998 | 2.028 | 1.969 | 1.999 | 2.028 | 0.952 | – |
| $\beta_3$ | 2.970 | 3.001 | 3.033 | 2.968 | 3.001 | 3.033 | 2.970 | 3.001 | 3.032 | 0.952 | – |
| $\omega$ | 0.457 | 0.498 | 0.537 | 0.456 | 0.497 | 0.537 | 0.457 | 0.497 | 0.537 | 0.952 | – |
| $\lambda$ | 3.806 | 5.143 | 7.036 | 3.577 | 4.828 | 6.536 | 3.688 | 4.982 | 6.740 | 0.947 | $1\times10^{-43}$ |

Table 3S: Point estimators, coverage proportions and Bayes factors: Linear regression model with residual errors simulated from a skew-logistic distribution, $n = 500$.

| Method | MLE | | | MAP | | | Median | | | Coverage | BF |
|---|---|---|---|---|---|---|---|---|---|---|---|
| | 5% | 50% | 95% | 5% | 50% | 95% | 5% | 50% | 95% | | |
| $\lambda = 0$ | | | | | | | | | | | |
| $\beta_1$ | 4.587 | 5.009 | 5.395 | 4.560 | 5.012 | 5.427 | 4.603 | 5.009 | 5.377 | 0.951 | – |
| $\beta_2$ | 1.891 | 1.999 | 2.109 | 1.890 | 1.999 | 2.112 | 1.890 | 1.998 | 2.108 | 0.950 | – |
| $\beta_3$ | 2.893 | 3.002 | 3.109 | 2.884 | 3.002 | 3.105 | 2.890 | 3.001 | 3.108 | 0.963 | – |
| $\omega$ | 0.403 | 0.525 | 0.732 | 0.429 | 0.557 | 0.735 | 0.446 | 0.584 | 0.768 | 0.859 | – |
| $\lambda$ | -1.064 | -0.032 | 1.096 | -0.907 | -0.021 | 0.861 | -0.984 | -0.026 | 0.989 | 0.951 | – |
| $\delta$ | 1.927 | 3.187 | 8.596 | 1.941 | 3.128 | 7.486 | 2.124 | 3.754 | 12.291 | 0.904 | 3.166 |
| $\lambda = 1$ | | | | | | | | | | | |
| $\beta_1$ | 4.780 | 4.996 | 5.262 | 4.753 | 4.983 | 5.269 | 4.789 | 5.007 | 5.273 | 0.960 | – |
| $\beta_2$ | 1.914 | 2.001 | 2.088 | 1.913 | 2.001 | 2.091 | 1.915 | 1.999 | 2.088 | 0.960 | – |
| $\beta_3$ | 2.917 | 2.999 | 3.086 | 2.911 | 2.997 | 3.089 | 2.917 | 2.998 | 3.088 | 0.959 | – |
| $\omega$ | 0.364 | 0.498 | 0.746 | 0.380 | 0.497 | 0.746 | 0.397 | 0.528 | 0.755 | 0.944 | – |
| $\lambda$ | 0.198 | 1.061 | 2.510 | 0.071 | 0.875 | 2.097 | 0.123 | 0.959 | 2.392 | 0.963 | – |
| $\delta$ | 1.918 | 3.162 | 10.156 | 1.917 | 3.021 | 6.954 | 2.095 | 3.633 | 11.724 | 0.914 | 0.966 |
| $\lambda = 2$ | | | | | | | | | | | |
| $\beta_1$ | 4.847 | 4.994 | 5.158 | 4.843 | 4.980 | 5.149 | 4.856 | 4.993 | 5.167 | 0.951 | – |
| $\beta_2$ | 1.933 | 2.000 | 2.074 | 1.928 | 2.001 | 2.076 | 1.930 | 2.000 | 2.074 | 0.950 | – |
| $\beta_3$ | 2.933 | 2.999 | 3.072 | 2.929 | 2.999 | 3.074 | 2.933 | 2.999 | 3.073 | 0.969 | – |
| $\omega$ | 0.361 | 0.507 | 0.717 | 0.351 | 0.509 | 0.728 | 0.377 | 0.522 | 0.725 | 0.949 | – |
| $\lambda$ | 1.001 | 2.262 | 5.496 | 0.782 | 1.867 | 3.979 | 0.912 | 2.091 | 4.776 | 0.950 | – |
| $\delta$ | 1.938 | 3.179 | 10.717 | 1.930 | 2.948 | 7.047 | 2.082 | 3.485 | 11.200 | 0.927 | 0.061 |
| $\lambda = 3$ | | | | | | | | | | | |
| $\beta_1$ | 4.883 | 4.991 | 5.106 | 4.880 | 4.985 | 5.104 | 4.893 | 4.990 | 5.110 | 0.952 | – |
| $\beta_2$ | 1.939 | 2.002 | 2.067 | 1.937 | 2.002 | 2.066 | 1.942 | 2.001 | 2.064 | 0.955 | – |
| $\beta_3$ | 2.943 | 3.000 | 3.061 | 2.942 | 2.999 | 3.062 | 2.944 | 2.999 | 3.059 | 0.966 | – |
| $\omega$ | 0.358 | 0.511 | 0.698 | 0.361 | 0.512 | 0.708 | 0.374 | 0.525 | 0.703 | 0.951 | – |
| $\lambda$ | 1.702 | 3.509 | 15.273 | 1.339 | 2.780 | 6.306 | 1.562 | 3.205 | 8.903 | 0.936 | – |
| $\delta$ | 1.947 | 3.212 | 13.439 | 1.923 | 2.968 | 6.882 | 2.101 | 3.494 | 11.393 | 0.921 | 0.015 |
| $\lambda = 4$ | | | | | | | | | | | |
| $\beta_1$ | 4.910 | 4.995 | 5.084 | 4.904 | 4.987 | 5.080 | 4.913 | 4.994 | 5.086 | 0.964 | – |
| $\beta_2$ | 1.945 | 2.002 | 2.062 | 1.942 | 2.002 | 2.060 | 1.946 | 2.002 | 2.059 | 0.949 | – |
| $\beta_3$ | 2.949 | 2.999 | 3.060 | 2.944 | 2.999 | 3.058 | 2.949 | 2.999 | 3.058 | 0.960 | – |
| $\omega$ | 0.368 | 0.511 | 0.694 | 0.371 | 0.516 | 0.708 | 0.383 | 0.526 | 0.699 | 0.957 | – |
| $\lambda$ | 2.214 | 4.847 | 29.796 | 1.805 | 3.645 | 8.605 | 2.096 | 4.349 | 13.213 | 0.957 | – |
| $\delta$ | 1.954 | 3.276 | 12.496 | 1.890 | 2.946 | 6.617 | 2.063 | 3.497 | 10.987 | 0.931 | 0.008 |
| $\lambda = 5$ | | | | | | | | | | | |
| $\beta_1$ | 4.925 | 4.997 | 5.072 | 4.919 | 4.991 | 5.067 | 4.925 | 4.994 | 5.069 | 0.961 | – |
| $\beta_2$ | 1.948 | 2.001 | 2.054 | 1.948 | 2.001 | 2.055 | 1.950 | 2.002 | 2.053 | 0.964 | – |
| $\beta_3$ | 2.951 | 3.000 | 3.055 | 2.948 | 3.000 | 3.052 | 2.951 | 3.000 | 3.050 | 0.978 | – |
| $\omega$ | 0.370 | 0.512 | 0.686 | 0.371 | 0.514 | 0.688 | 0.390 | 0.525 | 0.684 | 0.945 | – |
| $\lambda$ | 2.813 | 5.988 | 56.575 | 2.226 | 4.317 | 10.681 | 2.673 | 5.371 | 18.051 | 0.960 | – |
| $\delta$ | 1.954 | 3.218 | 12.533 | 1.873 | 2.950 | 6.665 | 2.052 | 3.498 | 10.688 | 0.930 | 0.007 |

Table 4S: Point estimators, coverage proportions and Bayes factors: Linear regression model with residual errors simulated from a skew-$t$ distribution with 3 degrees of freedom, $n = 100$.

| Method | MLE | | | MAP | | | Median | | | Coverage | BF |
|---|---|---|---|---|---|---|---|---|---|---|---|
| | 5% | 50% | 95% | 5% | 50% | 95% | 5% | 50% | 95% | | |
| $\lambda = 0$ | | | | | | | | | | | |
| $\beta_1$ | 4.776 | 4.997 | 5.235 | 4.749 | 4.997 | 5.244 | 4.780 | 4.998 | 5.227 | 0.964 | – |
| $\beta_2$ | 1.935 | 2.000 | 2.064 | 1.931 | 1.998 | 2.067 | 1.935 | 1.999 | 2.065 | 0.965 | – |
| $\beta_3$ | 2.937 | 3.000 | 3.066 | 2.934 | 3.000 | 3.070 | 2.936 | 3.000 | 3.067 | 0.953 | – |
| $\omega$ | 0.439 | 0.509 | 0.599 | 0.449 | 0.523 | 0.616 | 0.456 | 0.531 | 0.628 | 0.914 | – |
| $\lambda$ | -0.526 | 0.005 | 0.509 | -0.500 | 0.005 | 0.490 | -0.481 | 0.006 | 0.496 | 0.967 | – |
| $\delta$ | 2.252 | 3.040 | 4.944 | 2.249 | 3.022 | 4.803 | 2.369 | 3.195 | 5.480 | 0.927 | 5.582 |
| $\lambda = 1$ | | | | | | | | | | | |
| $\beta_1$ | 4.857 | 5.001 | 5.157 | 4.847 | 4.996 | 5.164 | 4.856 | 5.003 | 5.160 | 0.951 | – |
| $\beta_2$ | 1.950 | 1.998 | 2.052 | 1.949 | 2.000 | 2.055 | 1.951 | 1.998 | 2.052 | 0.960 | – |
| $\beta_3$ | 2.945 | 2.999 | 3.053 | 2.944 | 3.000 | 3.055 | 2.945 | 2.999 | 3.054 | 0.954 | – |
| $\omega$ | 0.408 | 0.499 | 0.639 | 0.407 | 0.494 | 0.633 | 0.420 | 0.507 | 0.641 | 0.946 | – |
| $\lambda$ | 0.461 | 1.024 | 1.768 | 0.401 | 0.962 | 1.720 | 0.430 | 0.989 | 1.759 | 0.954 | – |
| $\delta$ | 2.244 | 3.036 | 4.987 | 2.224 | 2.959 | 4.732 | 2.316 | 3.158 | 5.440 | 0.940 | 0.140 |
| $\lambda = 2$ | | | | | | | | | | | |
| $\beta_1$ | 4.914 | 4.999 | 5.088 | 4.911 | 4.997 | 5.091 | 4.914 | 5.000 | 5.091 | 0.945 | – |
| $\beta_2$ | 1.960 | 2.000 | 2.042 | 1.958 | 2.000 | 2.045 | 1.960 | 2.000 | 2.043 | 0.961 | – |
| $\beta_3$ | 2.955 | 2.998 | 3.042 | 2.952 | 2.999 | 3.042 | 2.955 | 2.999 | 3.042 | 0.955 | – |
| $\omega$ | 0.404 | 0.501 | 0.615 | 0.401 | 0.500 | 0.617 | 0.407 | 0.505 | 0.625 | 0.953 | – |
| $\lambda$ | 1.271 | 2.067 | 3.226 | 1.176 | 1.932 | 3.017 | 1.218 | 2.009 | 3.154 | 0.953 | – |
| $\delta$ | 2.232 | 3.050 | 4.992 | 2.194 | 2.944 | 4.575 | 2.288 | 3.152 | 5.214 | 0.949 | $8\times 10^{-9}$ |
| $\lambda = 3$ | | | | | | | | | | | |
| $\beta_1$ | 4.937 | 4.999 | 5.067 | 4.933 | 4.997 | 5.065 | 4.934 | 5.000 | 5.067 | 0.949 | – |
| $\beta_2$ | 1.966 | 2.000 | 2.038 | 1.963 | 2.000 | 2.038 | 1.965 | 2.000 | 2.037 | 0.953 | – |
| $\beta_3$ | 2.963 | 2.999 | 3.034 | 2.961 | 2.999 | 3.036 | 2.962 | 2.998 | 3.033 | 0.949 | – |
| $\omega$ | 0.410 | 0.502 | 0.604 | 0.406 | 0.502 | 0.614 | 0.416 | 0.507 | 0.612 | 0.956 | – |
| $\lambda$ | 2.030 | 3.169 | 5.009 | 1.825 | 2.866 | 4.571 | 1.949 | 3.042 | 4.807 | 0.948 | – |
| $\delta$ | 2.240 | 3.090 | 5.062 | 2.156 | 2.928 | 4.647 | 2.286 | 3.150 | 5.227 | 0.937 | $2\times 10^{-12}$ |
| $\lambda = 4$ | | | | | | | | | | | |
| $\beta_1$ | 4.950 | 5.000 | 5.056 | 4.948 | 4.997 | 5.058 | 4.950 | 5.000 | 5.058 | 0.953 | – |
| $\beta_2$ | 1.969 | 1.999 | 2.033 | 1.968 | 2.000 | 2.034 | 1.970 | 1.999 | 2.032 | 0.952 | – |
| $\beta_3$ | 2.967 | 2.999 | 3.031 | 2.966 | 2.998 | 3.032 | 2.968 | 2.999 | 3.032 | 0.959 | – |
| $\omega$ | 0.413 | 0.505 | 0.596 | 0.412 | 0.503 | 0.600 | 0.416 | 0.507 | 0.603 | 0.954 | – |
| $\lambda$ | 2.749 | 4.201 | 6.931 | 2.497 | 3.784 | 5.931 | 2.650 | 4.063 | 6.461 | 0.960 | – |
| $\delta$ | 2.243 | 3.081 | 4.863 | 2.187 | 2.965 | 4.488 | 2.289 | 3.147 | 5.086 | 0.946 | $6\times 10^{-12}$ |
| $\lambda = 5$ | | | | | | | | | | | |
| $\beta_1$ | 4.957 | 4.999 | 5.049 | 4.953 | 4.998 | 5.048 | 4.955 | 5.000 | 5.050 | 0.950 | – |
| $\beta_2$ | 1.973 | 2.000 | 2.029 | 1.971 | 1.999 | 2.029 | 1.972 | 1.999 | 2.029 | 0.955 | – |
| $\beta_3$ | 2.971 | 2.999 | 3.030 | 2.970 | 2.999 | 3.030 | 2.971 | 2.999 | 3.029 | 0.965 | – |
| $\omega$ | 0.412 | 0.504 | 0.591 | 0.413 | 0.504 | 0.599 | 0.415 | 0.507 | 0.598 | 0.945 | – |
| $\lambda$ | 3.315 | 5.228 | 9.096 | 2.939 | 4.638 | 7.558 | 3.161 | 4.977 | 8.352 | 0.955 | – |
| $\delta$ | 2.231 | 3.068 | 5.116 | 2.157 | 2.928 | 4.503 | 2.278 | 3.137 | 5.264 | 0.935 | $7\times 10^{-11}$ |

Table 5S: Point estimators, coverage proportions and Bayes factors: Linear regression model with residual errors simulated from a skew-$t$ distribution with 3 degrees of freedom, $n = 250$.

| Method | MLE | | | MAP | | | Median | | | Coverage | BF |
|---|---|---|---|---|---|---|---|---|---|---|---|
| | 5% | 50% | 95% | 5% | 50% | 95% | 5% | 50% | 95% | | |
| $\lambda = 0$ | | | | | | | | | | | |
| $\beta_1$ | 4.831 | 4.990 | 5.162 | 4.823 | 4.992 | 5.172 | 4.834 | 4.991 | 5.158 | 0.954 | – |
| $\beta_2$ | 1.951 | 1.999 | 2.042 | 1.951 | 1.999 | 2.045 | 1.951 | 2.000 | 2.043 | 0.954 | – |
| $\beta_3$ | 2.955 | 3.002 | 3.046 | 2.955 | 3.002 | 3.047 | 2.954 | 3.002 | 3.047 | 0.951 | – |
| $\omega$ | 0.455 | 0.506 | 0.565 | 0.460 | 0.513 | 0.575 | 0.464 | 0.517 | 0.579 | 0.918 | – |
| $\omega$ | -0.367 | 0.024 | 0.369 | -0.362 | 0.019 | 0.378 | -0.362 | 0.024 | 0.369 | 0.955 | – |
| $\delta$ | 2.414 | 3.060 | 4.103 | 2.381 | 3.044 | 4.040 | 2.463 | 3.133 | 4.269 | 0.940 | 8.122 |
| $\lambda = 1$ | | | | | | | | | | | |
| $\beta_1$ | 4.890 | 5.000 | 5.105 | 4.887 | 4.998 | 5.110 | 4.894 | 5.002 | 5.110 | 0.942 | – |
| $\beta_2$ | 1.960 | 2.000 | 2.035 | 1.958 | 2.000 | 2.037 | 1.960 | 1.999 | 2.035 | 0.947 | – |
| $\beta_3$ | 2.965 | 3.000 | 3.038 | 2.963 | 3.000 | 3.041 | 2.965 | 3.000 | 3.038 | 0.972 | – |
| $\omega$ | 0.433 | 0.501 | 0.594 | 0.430 | 0.499 | 0.595 | 0.437 | 0.505 | 0.595 | 0.955 | – |
| $\omega$ | 0.610 | 1.015 | 1.556 | 0.566 | 0.984 | 1.502 | 0.581 | 1.002 | 1.538 | 0.938 | – |
| $\delta$ | 2.416 | 3.058 | 4.121 | 2.415 | 3.020 | 4.025 | 2.475 | 3.118 | 4.212 | 0.937 | $5\times10^{-7}$ |
| $\lambda = 2$ | | | | | | | | | | | |
| $\beta_1$ | 4.938 | 4.999 | 5.066 | 4.931 | 4.998 | 5.065 | 4.937 | 5.001 | 5.069 | 0.941 | – |
| $\beta_2$ | 1.969 | 1.999 | 2.028 | 1.968 | 1.999 | 2.029 | 1.969 | 1.999 | 2.028 | 0.947 | – |
| $\beta_3$ | 2.972 | 3.000 | 3.030 | 2.971 | 3.001 | 3.031 | 2.972 | 3.000 | 3.030 | 0.965 | – |
| $\omega$ | 0.428 | 0.504 | 0.583 | 0.428 | 0.502 | 0.586 | 0.430 | 0.504 | 0.586 | 0.937 | – |
| $\omega$ | 1.445 | 2.032 | 2.877 | 1.387 | 1.962 | 2.775 | 1.428 | 2.001 | 2.838 | 0.943 | – |
| $\delta$ | 2.423 | 3.068 | 4.142 | 2.370 | 2.988 | 4.024 | 2.461 | 3.105 | 4.257 | 0.949 | $8\times10^{-35}$ |
| $\lambda = 3$ | | | | | | | | | | | |
| $\beta_1$ | 4.957 | 4.998 | 5.045 | 4.955 | 4.999 | 5.045 | 4.957 | 4.999 | 5.045 | 0.960 | – |
| $\beta_2$ | 1.973 | 1.999 | 2.024 | 1.973 | 1.999 | 2.025 | 1.973 | 1.999 | 2.024 | 0.952 | – |
| $\beta_3$ | 2.976 | 3.000 | 3.025 | 2.975 | 2.999 | 3.027 | 2.976 | 3.000 | 3.026 | 0.969 | – |
| $\omega$ | 0.434 | 0.504 | 0.573 | 0.433 | 0.504 | 0.575 | 0.435 | 0.506 | 0.573 | 0.957 | – |
| $\omega$ | 2.291 | 3.080 | 4.228 | 2.175 | 2.968 | 4.020 | 2.251 | 3.031 | 4.146 | 0.960 | – |
| $\delta$ | 2.432 | 3.056 | 4.075 | 2.376 | 2.999 | 3.942 | 2.456 | 3.081 | 4.127 | 0.950 | $1\times10^{-43}$ |
| $\lambda = 4$ | | | | | | | | | | | |
| $\beta_1$ | 4.964 | 4.999 | 5.033 | 4.963 | 4.999 | 5.034 | 4.964 | 4.999 | 5.035 | 0.958 | – |
| $\beta_2$ | 1.976 | 1.999 | 2.021 | 1.976 | 1.999 | 2.022 | 1.977 | 1.999 | 2.021 | 0.949 | – |
| $\beta_3$ | 2.978 | 3.000 | 3.022 | 2.978 | 3.000 | 3.023 | 2.979 | 3.000 | 3.022 | 0.958 | – |
| $\omega$ | 0.437 | 0.503 | 0.569 | 0.439 | 0.505 | 0.572 | 0.440 | 0.505 | 0.571 | 0.950 | – |
| $\omega$ | 3.052 | 4.112 | 5.798 | 2.911 | 3.915 | 5.533 | 2.986 | 4.018 | 5.666 | 0.956 | – |
| $\delta$ | 2.417 | 3.043 | 4.144 | 2.367 | 2.974 | 4.023 | 2.437 | 3.085 | 4.206 | 0.941 | $3\times10^{-41}$ |
| $\lambda = 5$ | | | | | | | | | | | |
| $\beta_1$ | 4.970 | 4.999 | 5.029 | 4.969 | 4.998 | 5.028 | 4.970 | 4.999 | 5.029 | 0.963 | – |
| $\beta_2$ | 1.979 | 1.999 | 2.019 | 1.978 | 1.999 | 2.020 | 1.978 | 1.999 | 2.019 | 0.956 | – |
| $\beta_3$ | 2.980 | 3.001 | 3.021 | 2.980 | 3.001 | 3.021 | 2.980 | 3.001 | 3.021 | 0.956 | – |
| $\omega$ | 0.443 | 0.503 | 0.563 | 0.445 | 0.502 | 0.567 | 0.445 | 0.505 | 0.566 | 0.949 | – |
| $\omega$ | 3.835 | 5.221 | 7.380 | 3.613 | 4.913 | 6.910 | 3.748 | 5.094 | 7.178 | 0.945 | – |
| $\delta$ | 2.412 | 3.066 | 4.086 | 2.366 | 2.985 | 3.962 | 2.442 | 3.094 | 4.159 | 0.953 | $3\times10^{-35}$ |

Table 6S: Point estimators, coverage proportions and Bayes factors: Linear regression model with residual errors simulated from a skew-$t$ distribution with 3 degrees of freedom, $n = 500$.

# 4 Simulation with prior (7)

For a description of the simulation scenario see Section 4 of the main paper.

| Method | MLE | | | MAP | | | Median | | | Coverage | BF |
|---|---|---|---|---|---|---|---|---|---|---|---|
| | 5% | 50% | 95% | 5% | 50% | 95% | 5% | 50% | 95% | | |
| $\lambda = 0$ | | | | | | | | | | | |
| $\beta_1$ | 4.477 | 4.991 | 5.535 | 4.617 | 4.999 | 5.404 | 4.802 | 4.997 | 5.247 | 0.999 | – |
| $\beta_2$ | 1.916 | 2.001 | 2.085 | 1.915 | 2.000 | 2.088 | 1.918 | 2.001 | 2.082 | 0.956 | – |
| $\beta_3$ | 2.916 | 2.997 | 3.088 | 2.913 | 2.997 | 3.088 | 2.914 | 2.997 | 3.087 | 0.950 | – |
| $\omega$ | 0.485 | 0.606 | 0.759 | 0.462 | 0.523 | 0.605 | 0.486 | 0.550 | 0.635 | 0.904 | – |
| $\lambda$ | -2.079 | -0.002 | 2.032 | -0.487 | -0.006 | 0.383 | -0.659 | 0.005 | 0.568 | 0.999 | 1.061 |
| $\lambda = 1$ | | | | | | | | | | | |
| $\beta_1$ | 4.790 | 5.010 | 5.318 | 4.870 | 5.249 | 5.474 | 4.969 | 5.242 | 5.400 | 0.915 | – |
| $\beta_2$ | 1.925 | 1.999 | 2.071 | 1.926 | 1.999 | 2.073 | 1.927 | 1.999 | 2.071 | 0.942 | – |
| $\beta_3$ | 2.927 | 3.001 | 3.065 | 2.925 | 3.002 | 3.070 | 2.927 | 3.002 | 3.065 | 0.961 | – |
| $\omega$ | 0.368 | 0.487 | 0.646 | 0.381 | 0.436 | 0.525 | 0.397 | 0.458 | 0.546 | 0.990 | – |
| $\lambda$ | 0.006 | 0.995 | 2.659 | -0.306 | 0.025 | 1.500 | -0.328 | 0.106 | 1.164 | 0.939 | 1.0321 |
| $\lambda = 2$ | | | | | | | | | | | |
| $\beta_1$ | 4.885 | 5.005 | 5.290 | 4.916 | 5.064 | 5.423 | 4.946 | 5.188 | 5.368 | 0.799 | – |
| $\beta_2$ | 1.944 | 2.000 | 2.056 | 1.941 | 2.000 | 2.056 | 1.945 | 2.000 | 2.055 | 0.963 | – |
| $\beta_3$ | 2.942 | 2.998 | 3.058 | 2.940 | 2.998 | 3.061 | 2.943 | 2.998 | 3.058 | 0.957 | – |
| $\omega$ | 0.342 | 0.491 | 0.603 | 0.321 | 0.380 | 0.558 | 0.340 | 0.410 | 0.548 | 0.838 | – |
| $\lambda$ | 0.240 | 2.116 | 5.181 | -0.195 | 0.230 | 2.779 | -0.007 | 0.672 | 2.980 | 0.783 | 0.749 |
| $\lambda = 3$ | | | | | | | | | | | |
| $\beta_1$ | 4.911 | 4.998 | 5.115 | 4.939 | 5.028 | 5.401 | 4.949 | 5.064 | 5.339 | 0.869 | – |
| $\beta_2$ | 1.951 | 2.000 | 2.051 | 1.949 | 2.001 | 2.048 | 1.950 | 2.001 | 2.048 | 0.960 | – |
| $\beta_3$ | 2.944 | 3.000 | 3.052 | 2.943 | 2.997 | 3.053 | 2.946 | 2.999 | 3.052 | 0.952 | – |
| $\omega$ | 0.394 | 0.494 | 0.595 | 0.307 | 0.449 | 0.560 | 0.331 | 0.446 | 0.553 | 0.863 | – |
| $\lambda$ | 1.512 | 3.289 | 10.371 | -0.121 | 1.992 | 3.904 | 0.180 | 1.918 | 4.416 | 0.835 | 0.273 |
| $\lambda = 4$ | | | | | | | | | | | |
| $\beta_1$ | 4.926 | 5.001 | 5.093 | 4.950 | 5.026 | 5.242 | 4.960 | 5.043 | 5.285 | 0.866 | – |
| $\beta_2$ | 1.953 | 1.999 | 2.046 | 1.953 | 2.000 | 2.048 | 1.954 | 2.000 | 2.046 | 0.951 | – |
| $\beta_3$ | 2.953 | 2.999 | 3.046 | 2.952 | 3.000 | 3.047 | 2.954 | 3.000 | 3.047 | 0.959 | – |
| $\omega$ | 0.407 | 0.495 | 0.585 | 0.305 | 0.460 | 0.553 | 0.335 | 0.456 | 0.555 | 0.883 | – |
| $\lambda$ | 2.176 | 4.378 | 115.504 | 0.014 | 2.550 | 4.974 | 0.413 | 2.739 | 5.896 | 0.849 | 0.105 |
| $\lambda = 5$ | | | | | | | | | | | |
| $\beta_1$ | 4.933 | 5.000 | 5.069 | 4.955 | 5.019 | 5.118 | 4.963 | 5.029 | 5.212 | 0.918 | – |
| $\beta_2$ | 1.957 | 1.998 | 2.042 | 1.957 | 1.998 | 2.043 | 1.958 | 1.998 | 2.042 | 0.964 | – |
| $\beta_3$ | 2.958 | 2.998 | 3.041 | 2.957 | 2.998 | 3.044 | 2.957 | 2.999 | 3.041 | 0.971 | – |
| $\omega$ | 0.416 | 0.497 | 0.576 | 0.308 | 0.468 | 0.549 | 0.350 | 0.466 | 0.552 | 0.916 | – |
| $\lambda$ | 2.882 | 5.844 | 135.799 | 0.119 | 3.197 | 6.105 | 0.906 | 3.579 | 7.721 | 0.895 | 0.027 |

Table 7S: Point estimators, coverage proportions and Bayes factors: Linear regression model with residual errors simulated from a skew-normal distribution, $n = 100$.

| Method | MLE | | | MAP | | | Median | | | Coverage | BF |
|---|---|---|---|---|---|---|---|---|---|---|---|
| | 5% | 50% | 95% | 5% | 50% | 95% | 5% | 50% | 95% | | |
| $\lambda = 0$ | | | | | | | | | | | |
| $\beta_1$ | 4.567 | 4.976 | 5.436 | 4.613 | 5.000 | 5.367 | 4.774 | 4.996 | 5.237 | 0.994 | – |
| $\beta_2$ | 1.945 | 2.002 | 2.052 | 1.945 | 2.001 | 2.053 | 1.946 | 2.001 | 2.052 | 0.954 | – |
| $\beta_3$ | 2.945 | 3.001 | 3.054 | 2.942 | 3.000 | 3.054 | 2.945 | 3.000 | 3.054 | 0.941 | – |
| $\omega$ | 0.498 | 0.584 | 0.688 | 0.479 | 0.519 | 0.577 | 0.497 | 0.539 | 0.603 | 0.894 | – |
| $\lambda$ | -1.363 | 0.038 | 1.452 | -0.912 | -0.002 | 0.929 | -0.609 | 0.001 | 0.612 | 0.993 | 1.166 |
| $\lambda = 1$ | | | | | | | | | | | |
| $\beta_1$ | 4.867 | 5.008 | 5.285 | 4.898 | 5.180 | 5.444 | 4.946 | 5.216 | 5.360 | 0.858 | – |
| $\beta_2$ | 1.959 | 1.999 | 2.041 | 1.955 | 1.999 | 2.043 | 1.958 | 1.999 | 2.041 | 0.944 | – |
| $\beta_3$ | 2.953 | 2.999 | 3.045 | 2.951 | 2.999 | 3.045 | 2.952 | 2.999 | 3.045 | 0.942 | – |
| $\omega$ | 0.396 | 0.493 | 0.589 | 0.394 | 0.430 | 0.554 | 0.410 | 0.450 | 0.540 | 0.968 | – |
| $\lambda$ | 0.017 | 0.994 | 1.873 | -0.281 | 0.108 | 1.556 | -0.206 | 0.190 | 1.370 | 0.862 | 1.095 |
| $\lambda = 2$ | | | | | | | | | | | |
| $\beta_1$ | 4.931 | 5.004 | 5.107 | 4.941 | 5.021 | 5.313 | 4.949 | 5.040 | 5.300 | 0.857 | – |
| $\beta_2$ | 1.964 | 2.001 | 2.036 | 1.961 | 2.001 | 2.037 | 1.963 | 2.001 | 2.034 | 0.951 | – |
| $\beta_3$ | 2.965 | 3.001 | 3.034 | 2.963 | 2.999 | 3.035 | 2.965 | 3.000 | 3.035 | 0.962 | – |
| $\omega$ | 0.420 | 0.494 | 0.559 | 0.340 | 0.474 | 0.552 | 0.362 | 0.466 | 0.547 | 0.858 | – |
| $\lambda$ | 1.145 | 2.000 | 3.017 | -0.055 | 1.665 | 2.603 | 0.230 | 1.585 | 2.631 | 0.842 | 0.205 |
| $\lambda = 3$ | | | | | | | | | | | |
| $\beta_1$ | 4.951 | 5.003 | 5.063 | 4.958 | 5.011 | 5.081 | 4.962 | 5.016 | 5.098 | 0.932 | – |
| $\beta_2$ | 1.970 | 2.000 | 2.031 | 1.969 | 2.000 | 2.032 | 1.970 | 2.000 | 2.031 | 0.964 | – |
| $\beta_3$ | 2.968 | 3.000 | 3.030 | 2.966 | 3.000 | 3.031 | 2.968 | 3.000 | 3.031 | 0.956 | – |
| $\omega$ | 0.439 | 0.496 | 0.552 | 0.418 | 0.485 | 0.542 | 0.417 | 0.485 | 0.542 | 0.936 | – |
| $\lambda$ | 2.080 | 3.067 | 4.780 | 1.633 | 2.583 | 3.845 | 1.600 | 2.685 | 4.076 | 0.917 | $5 \times 10^{-9}$ |
| $\lambda = 4$ | | | | | | | | | | | |
| $\beta_1$ | 4.957 | 4.999 | 5.047 | 4.962 | 5.007 | 5.058 | 4.965 | 5.009 | 5.059 | 0.953 | – |
| $\beta_2$ | 1.974 | 2.000 | 2.027 | 1.973 | 2.000 | 2.028 | 1.974 | 2.000 | 2.027 | 0.968 | – |
| $\beta_3$ | 2.972 | 3.000 | 3.025 | 2.971 | 3.000 | 3.025 | 2.972 | 3.000 | 3.026 | 0.974 | – |
| $\omega$ | 0.446 | 0.500 | 0.548 | 0.435 | 0.491 | 0.540 | 0.436 | 0.492 | 0.540 | 0.949 | – |
| $\lambda$ | 2.861 | 4.222 | 6.612 | 2.396 | 3.487 | 5.141 | 2.485 | 3.648 | 5.404 | 0.946 | $8 \times 10^{-13}$ |
| $\lambda = 5$ | | | | | | | | | | | |
| $\beta_1$ | 4.966 | 5.001 | 5.044 | 4.972 | 5.007 | 5.053 | 4.974 | 5.009 | 5.056 | 0.945 | – |
| $\beta_2$ | 1.973 | 2.000 | 2.025 | 1.973 | 2.000 | 2.025 | 1.974 | 2.000 | 2.025 | 0.959 | – |
| $\beta_3$ | 2.974 | 2.999 | 3.024 | 2.974 | 3.000 | 3.025 | 2.974 | 3.000 | 3.024 | 0.961 | – |
| $\omega$ | 0.452 | 0.497 | 0.546 | 0.443 | 0.489 | 0.535 | 0.443 | 0.490 | 0.538 | 0.956 | – |
| $\lambda$ | 3.481 | 5.202 | 8.767 | 2.888 | 4.263 | 6.408 | 3.048 | 4.488 | 6.859 | 0.937 | $7 \times 10^{-13}$ |

Table 8S: Point estimators, coverage proportions and Bayes factors: Linear regression model with residual errors simulated from a skew-normal distribution, $n = 250$.

| Method | MLE | | | MAP | | | Median | | | Coverage | BF |
|---|---|---|---|---|---|---|---|---|---|---|---|
| | 5% | 50% | 95% | 5% | 50% | 95% | 5% | 50% | 95% | | |
| $\lambda = 0$ | | | | | | | | | | | |
| $\beta_1$ | 4.617 | 4.994 | 5.379 | 4.642 | 5.004 | 5.354 | 4.776 | 4.997 | 5.228 | 0.994 | – |
| $\beta_2$ | 1.962 | 2.000 | 2.039 | 1.960 | 1.999 | 2.040 | 1.961 | 2.000 | 2.040 | 0.939 | – |
| $\beta_3$ | 2.963 | 3.001 | 3.037 | 2.960 | 3.001 | 3.041 | 2.962 | 3.001 | 3.037 | 0.948 | – |
| $\omega$ | 0.501 | 0.568 | 0.648 | 0.485 | 0.514 | 0.555 | 0.499 | 0.532 | 0.587 | 0.884 | – |
| $\lambda$ | -1.142 | 0.003 | 1.140 | -0.952 | -0.010 | 0.954 | -0.590 | 0.005 | 0.602 | 0.994 | 1.276 |
| $\lambda = 1$ | | | | | | | | | | | |
| $\beta_1$ | 4.900 | 5.000 | 5.267 | 4.910 | 5.067 | 5.400 | 4.943 | 5.178 | 5.319 | 0.826 | – |
| $\beta_2$ | 1.970 | 2.001 | 2.030 | 1.968 | 2.000 | 2.031 | 1.969 | 2.001 | 2.030 | 0.958 | – |
| $\beta_3$ | 2.970 | 2.999 | 3.031 | 2.969 | 2.998 | 3.030 | 2.969 | 2.999 | 3.030 | 0.951 | – |
| $\omega$ | 0.405 | 0.499 | 0.562 | 0.402 | 0.429 | 0.551 | 0.414 | 0.448 | 0.535 | 0.927 | – |
| $\lambda$ | 0.047 | 1.014 | 1.561 | -0.307 | 0.360 | 1.408 | -0.105 | 0.327 | 1.314 | 0.813 | 1.075 |
| $\lambda = 2$ | | | | | | | | | | | |
| $\beta_1$ | 4.947 | 5.003 | 5.065 | 4.952 | 5.010 | 5.081 | 4.956 | 5.016 | 5.114 | 0.916 | – |
| $\beta_2$ | 1.976 | 2.001 | 2.025 | 1.976 | 2.001 | 2.026 | 1.976 | 2.001 | 2.025 | 0.964 | – |
| $\beta_3$ | 2.975 | 3.000 | 3.024 | 2.974 | 3.000 | 3.026 | 2.975 | 3.000 | 3.024 | 0.955 | – |
| $\omega$ | 0.445 | 0.497 | 0.546 | 0.419 | 0.490 | 0.542 | 0.417 | 0.488 | 0.540 | 0.910 | – |
| $\lambda$ | 1.436 | 1.983 | 2.682 | 1.227 | 1.826 | 2.488 | 1.070 | 1.837 | 2.530 | 0.913 | $1\times10^{-17}$ |
| $\lambda = 3$ | | | | | | | | | | | |
| $\beta_1$ | 4.962 | 5.000 | 5.042 | 4.963 | 5.004 | 5.049 | 4.967 | 5.006 | 5.051 | 0.945 | – |
| $\beta_2$ | 1.977 | 1.999 | 2.020 | 1.977 | 2.000 | 2.020 | 1.978 | 1.999 | 2.020 | 0.957 | – |
| $\beta_3$ | 2.979 | 3.000 | 3.021 | 2.978 | 3.000 | 3.022 | 2.979 | 3.000 | 3.021 | 0.962 | – |
| $\omega$ | 0.461 | 0.500 | 0.537 | 0.452 | 0.494 | 0.535 | 0.454 | 0.495 | 0.535 | 0.944 | – |
| $\lambda$ | 2.325 | 3.050 | 4.081 | 2.113 | 2.810 | 3.691 | 2.162 | 2.876 | 3.791 | 0.938 | $5\times10^{-53}$ |
| $\lambda = 4$ | | | | | | | | | | | |
| $\beta_1$ | 4.968 | 5.001 | 5.033 | 4.970 | 5.004 | 5.038 | 4.972 | 5.005 | 5.039 | 0.945 | – |
| $\beta_2$ | 1.982 | 2.000 | 2.020 | 1.982 | 2.001 | 2.020 | 1.982 | 2.000 | 2.020 | 0.972 | – |
| $\beta_3$ | 2.979 | 3.000 | 3.019 | 2.979 | 3.000 | 3.020 | 2.980 | 3.000 | 3.019 | 0.949 | – |
| $\omega$ | 0.461 | 0.499 | 0.532 | 0.457 | 0.494 | 0.528 | 0.457 | 0.495 | 0.528 | 0.952 | – |
| $\lambda$ | 3.064 | 4.049 | 5.470 | 2.826 | 3.700 | 4.981 | 2.879 | 3.796 | 5.074 | 0.934 | $1\times10^{-52}$ |
| $\lambda = 5$ | | | | | | | | | | | |
| $\beta_1$ | 4.976 | 5.000 | 5.028 | 4.978 | 5.003 | 5.033 | 4.979 | 5.004 | 5.033 | 0.956 | – |
| $\beta_2$ | 1.981 | 2.000 | 2.019 | 1.981 | 2.000 | 2.019 | 1.981 | 2.000 | 2.018 | 0.951 | – |
| $\beta_3$ | 2.982 | 3.000 | 3.017 | 2.981 | 3.000 | 3.018 | 2.982 | 3.000 | 3.017 | 0.959 | – |
| $\omega$ | 0.464 | 0.497 | 0.532 | 0.460 | 0.494 | 0.528 | 0.461 | 0.494 | 0.529 | 0.943 | – |
| $\lambda$ | 3.877 | 5.086 | 7.157 | 3.514 | 4.626 | 6.353 | 3.601 | 4.729 | 6.475 | 0.947 | $4\times10^{-48}$ |

Table 9S: Point estimators, coverage proportions and Bayes factors: Linear regression model with residual errors simulated from a skew-normal distribution, $n = 500$.

# References

[1] A. Azzalini and A. Capitanio. Distributions generated by perturbation of symmetry with emphasis on a multivariate skew-t distribution. *Journal of Royal Statistical Society, Series B*, 65:367–389, 2003.

[2] A. Azzalini and A. Dalla-Valle. The multivariate skew-normal distribution. *Biometrika*, 83:715–726, 1996.

[3] G.B. Dantzig and M.N. Thapa. *Linear programming 1: Introduction*. Springer Verlag, 1997.

[4] C. Fernández and M.F.J. Steel. Multivariate Student-$t$ regression models: Pitfalls and inference. *Biometrika*, 86:153–167, 1999.

[5] C. Fernández and M.F.J. Steel. Bayesian regression analysis with scale mixtures of normals. *Econometric Theory*, 80:80–101, 2000.

[6] C. Vallejos and M.F.J. Steel. Objective Bayesian survival analysis using shape mixtures of log-normal distributions. *Journal of the American Statistical Association*, 110:697–710, 2015.

[7] J. Wang, J. Boyer, and M.G. Genton. A skew symmetric representation of multivariate distributions. *Statistica Sinica*, 14:1259–1270, 2004.